\documentclass[12pt]{iopart}
\usepackage{graphicx}
\usepackage{color}
\expandafter\let\csname equation*\endcsname\relax
\expandafter\let\csname endequation*\endcsname\relax
\usepackage[]{amsmath}
\usepackage{amssymb}
\usepackage{hyperref}
\usepackage{enumerate}

\newcommand{\RR}{\mathbb{R}}
\newcommand{\EE}{\mathbb{E}}
\newcommand{\PP}{\mathbb{P}}

\newcommand{\Proj}{P}
\newcommand{\psin}{z}

\begin{document}

\title{Non-equilibrium transitions in multiscale systems with a
  bifurcating slow manifold}

\author{Tobias Grafke$^1$ and Eric Vanden-Eijnden$^1$}
\address{$^1$ Courant Institute, 
  New York University, 
  251 Mercer Street, New York, 
  NY 10012, USA}
\eads{\mailto{grafke@cims.nyu.edu}, \mailto{eve2@cims.nyu.edu}}

\date{\today}

\begin{abstract}
  Noise-induced transitions between metastable fixed points in systems
  evolving on multiple time scales are analyzed in situations where
  the time scale separation gives rise to a slow manifold with
  bifurcation. This analysis is performed within the realm of large
  deviation theory. It is shown that these non-equilibrium transitions
  make use of a reaction channel created by the bifurcation structure
  of the slow manifold, leading to vastly increased transition
  rates. Several examples are used to illustrate these findings,
  including an insect outbreak model, a system modeling phase
  separation in the presence of evaporation, and a system modeling
  transitions in active matter self-assembly. The last example
  involves a spatially extended system modeled by a stochastic partial
  differential equation.
\end{abstract}

%\tableofcontents

\maketitle

\section{Introduction}

Systems in nature are commonly found to evolve on multiple time
scales. One typically distinguishes \emph{slow variables} $x\in\RR^n$,
that change on a time scale slow compared to the \emph{fast variables}
$y\in\RR^m$. In a traditional setup
\cite{berglund-gentz:2006,kuehn2011mathematical}, these systems can be
written as ODEs of the form
\begin{equation}
  \label{eq:slowfastfg}
  \begin{aligned}
    \dot x &= \alpha u(x,y)\\
    \dot y &= v(x,y)\,.
  \end{aligned}
\end{equation}
where the parameter $\alpha$ measures how fast the variables $y$
evolve compared to $x$.  In the limit $\alpha\to0$ when the time scale
separation becomes infinite, some concepts are commonly defined for
slow-fast systems of the above form, in particular:
\begin{itemize}
\item the system may converge to a \emph{slow manifold} $\mathcal{M}$
  where $v(x,y)=0$ on the fast scale, so that the dynamics, after a
  short transient, are reduced to a movement closely along this
  manifold, and;
\item changes in stability along the slow manifold may lead to
  \emph{bifurcation}, with the slow variables playing the role of
  \emph{control parameters}.
\end{itemize}

On top of the deterministic dynamics of~\eqref{eq:slowfastfg}, both
the slow and the fast variables in many realistic systems are subject
to small intrinsic stochastic fluctuations, for example due to thermal
noise, or the discreteness of an underlying microscopic model. These
fluctuations introduce another time scale: the system may be confined
in a metastable region of the slow-fast dynamics for long periods of
time and only rarely switches between different such
regions. Specifically, in the case where the deterministic dynamics
of~\eqref{eq:slowfastfg} permit multiple locally stable fixed points
$(x_i,y_i)$ such that $u(x_i,y_i)=v(x_i,y_i)=0$, small fluctuations of
order $\mathcal{O}(\epsilon)$ may push the system from the vicinity of
one stable fixed-point to another. On time scales larger than the
Kramers' time, which is typically $\mathcal{O}(\exp(\epsilon^{-1}))$,
these transitions become almost certain, and are captured precisely by
\emph{large deviation theory} (LDT)~\cite{freidlin-wentzell:1998}. On
these time scales, the dynamics can then be effectively reduced to
that of a Markov jump process on the metastable
states~\cite{freidlin-wentzell:1998,berglund-gentz:2006}. In the
present paper, our aim is to analyze how the slow-fast nature of the
dynamics and its associated bifurcation structure influence the
pathways of the noise-induced transitions between metastable
fixed-points. We will focus mostly on situations where the Kramers'
time is the longest time scale of the system. The presence of three
time scales in interaction with the bifurcation structure quickly
leads to a rich set of possible phenomena. The objective of this paper
is to identify common patterns in these situations and apply them to
specific applications as a first step towards an exhaustive
theory. Our main contribution is to elucidate the role of the
bifurcation structure of the slow-fast time scale separation on
metastability. Interesting phenomena also occur in regimes where the
Kramers' time is between slow and fast time scales
\cite{berglund2002metastability,berglund-gentz:2006,kuehn2011mathematical},
like for example in the context of coherent or stochastic resonance phenomena
\cite{pikovsky-kurths:1997,freidlin2001stable,freidlin2001stochastic,muratov2005self,
  deville-vanden-eijnden-muratov:2005}. Similarly, one can consider
the case where the large deviation smallness parameter $\epsilon$ is
the time scale separation $\alpha$ itself \cite{freidlin:1978,
  kifer:1992, veretennikov:2000,bouchet2016large}. These situations
are not considered here.

The remainder of this paper is organized as follows.  In
Section~\ref{sec:set-up-intuition} we present the type of fast slow
system we will study and discuss their slow manifold and bifurcation
structures. We also introduce the effect of fluctuations on the
dynamics, leading to noise induced transitions between metastable
fixed points captured by LDT. Some generic cases are analyzed
explicitly in Section~\ref{sec:generic-examples}. In
Section~\ref{sec:finite-dimens-appl}, we illustrate the theory for
several finite dimensional applications that cover different
bifurcation structures, including a model for insect outbreak in
Section~\ref{sec:insect-outbreak} and a model for phase separation in
the presence of evaporation in Section~\ref{sec:phase-separ-with}. The
generalization to infinite dimensional applications is presented in
Section~\ref{sec:infin-dimens-appl}, where the theory is applied to a
stochastic partial differential equation modeling motility induced
phase separation for motile microorganisms. The appendix covers
details concerning the numerical computation of the most likely
transition pathway in~\ref{sec:sgmam} and of the bifurcation structure
in~\ref{sec:slowmanifold}.

\section{Problem set-up and phenomenology}
\label{sec:set-up-intuition}

\subsection{Bifurcation and control parameter}

We are going to use a generalized form of the
system~\eqref{eq:slowfastfg} that describes the evolution of a single
variable $z\in\RR^d$ obeying
\begin{equation}
  \label{eq:slowfast_det}
  \dot z = f(z) + \alpha g(z)
\end{equation}
where $f(z)$ describes the fast dynamics and $g(z)$ the slow dynamics,
and the time scale separation is measured by $\alpha\ll1$. All systems
of type~\eqref{eq:slowfastfg} can be brought into this form by taking
$z=(x,y)$, but in general for equation~\eqref{eq:slowfast_det}, it is
not clear \emph{a priori} what the slow and fast variables are (as
they are not necessarily isolated components of $z$). We call the set
\begin{equation}
  \mathcal{M} = \{v\in\RR^d|f(v)=0\}
\end{equation}
the \emph{slow manifold} of the system~\eqref{eq:slowfast_det}, to
which the system converges via the \emph{fast dynamics}
$\dot z = f(z)$ when we take $\alpha\ll0$. In other words, the fixed
points of the fast dynamics make up the slow manifold
$\mathcal{M}$. The slow manifold is comprised of \emph{stable}
branches, i.e. sets of points $v\in\mathcal{M}$ for which
$\nabla f(v)$ has eigenvalues with negative real part only, as well as
\emph{unstable} branches, where at least one eigenvalue has a positive
real part.

We call \emph{slow variables} all quantities left \emph{invariant} on
the fast time scale, i.e. slow variables are \emph{conserved
  quantities} of the fast dynamics. Dynamics on the slow time scale
are described by the \emph{reduced dynamics} defined by the projection
of the slow dynamics into the tangent space of the slow manifold.

Furthermore, we can identify all points on the slow manifold by the
associated values of the slow variables. In other words, the slow
manifold is foliated by the slow variables. As a consequence, it makes
sense to use the slow variables as \emph{control parameters} $\mu$ for
a bifurcation analysis. Varying the control parameter, branches of the
slow manifold appear or disappear, split or merge. The points
$v\in\mathcal{M}$ where this happens are called \emph{bifurcation
  points}, which are equivalent to points where the stability of the
slow manifold changes.

Note that all above definitions solely reference the fast dynamics
$f(z)$, regardless of the choice of $g(z)$. For small but finite time
scale separation $\alpha$, the picture is necessarily perturbed: The
reduced dynamics do not track the slow manifold exactly, changes of
stability do not occur exactly at bifurcation point, etc.

\subsection{Noise-induced transitions}

We now want to introduce stochastic fluctuations to the slow-fast
system~\eqref{eq:slowfast_det}, by taking
\begin{equation}
  \label{eq:slowfast}
  dZ = (f(Z) + \alpha g(Z))\,dt + \sqrt{\epsilon} \sigma dW\,.
\end{equation}
Here, $\epsilon$ is a parameter that will be considered small (as we
are interested in the low noise regime), $\sigma\in\RR^{d\times d}$ is
the noise correlation matrix and $W$ is a Wiener process (Brownian
motion). The noise term is to be understood as the combined effect of
all fluctuations of the system, and is assumed to be Gaussian. In
principle, if we assume slow and fast dynamics to be induced by two
separate physical mechanisms, each might be attained with its own
intrinsic fluctuations, acting on different time scales, with
different correlations, possibly degenerate and possibly
non-Gaussian. We will consider situations of this type later, but for
time being focus on the cases where the effect of the noise can be
combined as in equation~\eqref{eq:slowfast} with $\sigma\sigma^T$
invertible (i.e. the diffusion~\eqref{eq:slowfast} is elliptic) and
$\sigma$ is independent of $\alpha$. In the small noise limit,
$\epsilon\to0$, we will start the discussion with systems that have
only two distinct deterministically stable fixed points $z_A$ and
$z_B$. The same arguments can be applied to any pair of an arbitrary
higher number of stable fixed points (for more generic attractors,
such as limit cycles, the situation becomes more
complicated). Limiting ourselves to two stable fixed points $z_A$ and
$z_B$, we are interested in the noise-induced transitions between
them. For deterministic dynamics, $\epsilon=0$, each of the stable
fixed points is surrounded by its respective basin of attraction,
\begin{equation}
  \mathcal{B}_{i}=\left\{z\in\RR^d\,\Big|\,
    \lim_{t\to\infty}\left|z(t)-z_{i}\right|=0 \text{ for } \epsilon=0
  \text{ if } z(0)=z\right\}\,,\qquad i=A,B\,,
\end{equation}
so that all deterministic trajectories that are initially located in
$\mathcal{B}_{i}$ will converge arbitrarily close to $z_{i}$ for
large times. The two basins are separated by a separatrix, which, in
the simplest case, contains a single saddle point $z_S$, i.e. a fixed
point where one eigenvalue of $\nabla(f(z_S)+\alpha g(z_S))$ is
negative, while all others have a positive real part.

Adding fluctuations, the generic picture of the low-noise limit
implies that, after an initial transient, the system spends most of
its time close to one of the stable fixed points. Only rarely
excursions occur that push the state over the separatrix into the other
basin of attraction, where it will deterministically converge close to
the other deterministically stable fixed point. In the limit
$\epsilon\to0$ these transitions can be described precisely by
Freidlin-Wentzell theory of large deviations
\cite{freidlin-wentzell:1998}: To compute the most probable transition
pathway for a transition time $T$, we find the trajectory $\psin$ whose
Freidlin-Wentzell action or \emph{rate function}
\begin{equation}
  \label{eq:action}
  S^\alpha_T(\psin) = \frac12\int_0^T \left|\sigma^{-1} \left(\dot\psin -
      f(\psin)
      -\alpha g(\psin)\right)\right|^2\,dt
\end{equation}
is minimal.
% -- here we have assumed for simplicity that $\sigma$ is
% invertible, i.e. the diffusion~\eqref{eq:slowfast} is elliptic. 
Since
in general $T$ is not prescribed and we want to find the most probable
transition for an arbitrary transition time, we look at the double
minimization
\begin{equation}
  \label{eq:actionmin}
  \inf_{T>0} \inf_{\psin} S^\alpha_T(\psin)\,.
\end{equation}
Typically, since we are starting from (and ending at) a stable fixed
point, the minimum will be achieved when $T\to\infty$, meaning that
\eqref{eq:actionmin} has no minimizer unless we reparametrize it in
time: understanding the minimizer in this general sense, we will
denote it by $\psin^\star$, and refer to it as the \emph{maximum
  likelihood transition pathway} (MLP) or \emph{instanton}.

In the gradient case, and with $\sigma = \sqrt{2} \, \text{\it Id}$, the
system~\eqref{eq:slowfast} is in detailed balance, and there exists a
potential or free energy $U_\alpha:\RR^d\to\RR$ such that $f(z)+\alpha
g(z) = -\nabla U_\alpha(z)\,\forall\,z\in\RR^d$. This assumption
significantly simplifies the large deviation computation. In
particular, the `uphill'' portion of the transition (i.e. the portion
where fluctuations are necessary to overcome the deterministic
dynamics), fulfills
\begin{equation}
  \frac14\int_{0}^{T} |\dot\psin + \nabla U_\alpha(\psin)|\,dt = 
  \frac14 \int_{0}^{T}|\dot\psin-\nabla U_\alpha(\psin)|\,dt +
  (U_\alpha(\psin(T)) - U_\alpha(\psin(0)))\,,
\end{equation}
which is minimized by $\dot\psin^\star = \nabla
U_\alpha(\psin^\star)$. The rate function is therefore equal to the
\emph{barrier height} $U_\alpha(\psin(T)) - U_\alpha(\psin(0))$ along
the MLP. As a direct consequence, the MLP crosses the separatrix at
the saddle point $z_S$, where the barrier is lowest, i.e.~the barrier
height corresponds to the potential difference between the fixed point
$z_A$ and the saddle point $z_S$. Furthermore, since for all $t$,
$\dot\psin^\star(t)$ is parallel to $\nabla U_\alpha(\psin^\star(t))$, the MLP
coincides with the \emph{heteroclinic orbits} (HO) connecting $z_S$ to
$z_A$ and $z_B$. Note also that this implies that forward and backward
transitions follow the same path in reverse. Indeed the ``uphill''
portion obeys exactly the time-reversed dynamics of the ``downhill''
portion, which is a direct consequence of the detailed balance
property that holds in the equilibrium case.

In the non-equilibrium case, where detailed balance is broken and the
HO no longer coincides with the MLP, in general we have to resort to
numerical minimization of the action functional~\eqref{eq:action} to
obtain the MLP. In the setup lined out above, though, it remains
correct that once the MLP crosses the separatrix, it will obey the
deterministic dynamics, and it will cross the separatrix at the saddle
point $z_S$. On the other hand, it is no longer true that forward and
backward transitions are simply time-reversed, and detailed balance is
broken. Instead, forward and backward transitions typically form a
``figure-eight'' shape, where the parts $z_A \to z_S$ and
$z_S \to z_A$ (resp. $z_B\to z_S$ and $z_S\to z_B$) occur along
different paths, but both forward and backward transition meet at
$z_S$. In this case, there still is merit in computing the
heteroclinic orbit even in the non-equilibrium setup, as it will yield
the saddle point and half of each transition. (This intuition breaks
down as soon as there are several saddle points, or a more complicated
fixed-point structure on the separatrix. It is then possible that
forward and backward transition share none but their initial and final
points.)

\subsection{Transitions for large time scale separation}

The interplay between the stochastic fluctuations and the time scale
separation warrants a separate discussion. In the case of large time
scale separation, $\alpha\ll1$, the stable fixed points necessarily
lie close to the slow manifold $\mathcal{M}$, since either $f(z_{i})$
and $g(z_{i})$ individually vanish for $i=A,B$, and therefore
$z_{i}\in\mathcal{M}$, or they cancel each other, in which case
$\min_{v\in\mathcal{M}}|z_{i}-v| = \mathcal{O}(\alpha)$. The same is
true for all fixed points of the deterministic dynamics, including the
saddle $z_S$. This is of particular importance, since it implies that
in the limit $\alpha\to0$ (a) the separatrix (locally) coincides with
the slow manifold, and (b) the slow manifold is unstable around
$z_S$. The situation becomes particularly interesting if the two
deterministically stable fixed points $z_A, z_B$ are \emph{connected}
by the slow manifold in the limit $\alpha\to0$. The intuition then is
that the slow manifold opens a reactive channel for the transition,
which will be used by the maximum likelihood transition pathway. This
is plausible, since the reduced dynamics along the slow manifold are
of order $\mathcal{O}(\alpha)$ and therefore are much easier to
overcome by random fluctuations than the $\mathcal{O}(1)$ dynamics
elsewhere.

To be more precise, we can rescale the time variable in the
action~\eqref{eq:action} to the slow time scale $\tau=\alpha t$ to
obtain
\begin{equation}
  \alpha^{-1} S^\alpha_{T'/\alpha}(\psin) = \frac12\int_0^{T'} \left|\sigma^{-1} \left(\dot\psin -
      \alpha^{-1} f(\psin)
      -g(\psin)\right)\right|^2\,d\tau
\end{equation}
where $T' = \alpha T$.  Considering the limit $\alpha\to0$ at $T'$
fixed, the contribution of the fast dynamics becomes unbounded as soon
as the trajectory leaves the slow manifold, where $f=0$. Indeed given
a trajectory~$\psin$
\begin{equation}
  \label{eq:3}
  \lim_{\alpha\to0} \alpha^{-1} S^\alpha_{T'/\alpha}(\psin) < \infty\quad\text{if
    and only if $\psin \in \mathcal M$}%, but}\quad \lim_{\alpha\to0} \alpha^{-1}_T(\psin)=\infty\quad\text{if $\psin\notin\mathcal M$.}
\end{equation}
From this statement it follows that any transition of the
system~(\ref{eq:slowfast}) in the large deviation regime between
metastable fixed points that are connected by the slow manifold
$\mathcal M$ will reach a bifurcation point $z_X$ in the limit
$\alpha\to0$. This portion of the transition will be the only one that
contributes to the action, while the motion after $z_X$ is
deterministic.

Note though that the mode of descend from the bifurcation point $z_X$
onwards is nontrivial. Since the separatrix is unstable, the
transition trajectory may leave it at any point up to the saddle, and
deterministically relax into the opposing fixed point. The action
attained along all of these trajectories is zero, but they each carry
some probability current. A more refined estimate of the action than
our approach above is necessary to analyze their relative weight. Note
though that the leading scaling of the transition probability itself
is unaffected by these complications. A more detailed analysis of
similar setups is performed in the literature~\cite{maier-stein:1997,
  bouchet-touchette:2012}. For any finite but small value of $\alpha$,
on the other hand, the separatrix has to be crossed precisely at the
saddle $z_S$, but due to the scaling of the action~\eqref{eq:3} the
slow manifold is still used as reactive channel. Since the slow
manifold is locally stable around $z_{A}$ and $z_{B}$, and locally
unstable around $z_S$, the MLP must also closely pass a bifurcation
point $z_X$, where $\mathcal{M}$ changes stability. This implies the
counter-intuitive fact that the MLP approaches the separatrix already
at $z_X$, which is far from the saddle $z_S$, and then ``skirts'' the
separatrix for an extended time quasi-deterministically into $z_S$. We
will confirm this intuition in a number of examples below.

The setup lined out above, where the stable fixed points of a
slow-fast system are connected by the slow manifold, turns out to be
quite ubiquitous in nature, with applications in physics (active
Brownian particles, active matter), chemistry (reaction-diffusion),
and biology (motile microorganisms, chemotaxis, predator prey
models). In the sequel, we will study several examples arising in
these contexts. In particular we will discuss also an infinite
dimensional example (that is, including a spatial dimension): Active
particle phase separation, where the fast phase separation term is
combined with a slow destabilizing term, a model also applicable to
motile microorganisms with a fast propulsion and slow growth term.

\subsection{The case with slow-fast noises}

It is also useful to consider situations in which the fluctuations on
the slow and fast variables act on their respective time scales, that
is, models of the type
\begin{equation}
  \label{eq:slowfastfgSDE}
  \begin{aligned}
    d X &= \alpha u(X,Y)dt + \sqrt{\alpha\epsilon} \sigma_x dW_x\\
    dY &= v(X,Y) dt +\sqrt{\epsilon} \sigma_y dW_y\,.
  \end{aligned}
\end{equation}
where $W_x$ and $W_y$ are independent and both $\sigma_x\sigma_x^T$
and $\sigma_y\sigma_y^T$ are invertible and independent of $\alpha$. In this case, the
Freidlin-Wentzell action
reads
\begin{equation}
  \label{eq:3a}
  S^\alpha_T(x,y) = \frac12\int_0^T \left(\alpha^{-1}\left|\sigma_x^{-1} \left(\dot x -
      \alpha u(x,y)\right)\right|^2 +  \left|\sigma_y^{-1} \left(\dot y -
      v(x,y)\right)\right|^2\right)\,dt
\end{equation}
On the slow time scale $\tau=\alpha t$, this action can be equivalently
written as
\begin{equation}
  \label{eq:3b}
  S^\alpha_{T'/\alpha}(x,y) = \frac12\int_0^{T'} \left(\left|\sigma_x^{-1} \left(dx/d\tau -
      u(x,y)\right)\right|^2 +  \left|\sigma_y^{-1} \left(dy/d\tau -
      \alpha^{-1} v(x,y)\right)\right|^2\right)\,d\tau
\end{equation}
These two expressions indicate that the action remains finite as
$\alpha\to0$ if it is evaluated on paths that either follow the slow
manifold where $v(x,y) =0$ on the slow time scale, or take shortcuts at
$x=cst$ (so that $\dot x=0$) on the fast time scale. This typically
provides different possible scenarios for the transition, among which
the most likely one (i.e. the one minimizing the action) will depend
on the relative amplitude of $\sigma_x$ and $\sigma_y$. This will be
illustrated via examples below.

Note that systems generalizing~\eqref{eq:slowfastfgSDE} are consistent
with examples in which detailed balance is broken in a very specific
way. Namely, situations where both fast and slow dynamics, taken by
themselves, are in detailed balance with respect to their
corresponding fluctuations, but in combination they are not. This
setup is very natural: It occurs when both the slow and the fast term
are modeling physical processes that are in detailed balance
individually. In this case, the dynamics can be written as
\begin{equation}
  \label{eq:gradient-sum}
  dZ = -M_1\nabla E_1(Z)\,dt - \alpha M_2 E_2(Z)\,dt + 
  \sqrt{\epsilon_1} M_1^{1/2}\,dW_1 + \sqrt{\alpha\epsilon_2} M_2^{1/2}\,dW_2
\end{equation}
with free energies $E_i(z)$ and mobilities $M_i$, $i=1,2$. If the mobilities
are \emph{incompatible}, i.e. if one cannot find a single mobility $M$
such that the sum of the two gradient systems can be written as a
single gradient with mobility $M$, then the system can no longer be in
detailed balance, regardless of the choice of the noise
correlation. The complete system then describes the competing
influence of two distinct physical processes acting on different time
scales. The incompatibility arises, for example, if one of the
processes is conservative, (i.e. diffusion, advection, etc), while the
other is not (i.e. evaporation, reaction, birth/death, etc). In this
case, the conserved quantity also defines the slow variable and
therefore the control parameter $\mu$ for the bifurcation. For example
for conservative fast dynamics the control parameter is the spatial
mean of the considered field variable. Both the finite dimensional
phase separation/evaporation discussed in
section~\ref{sec:phase-separ-with} and the infinite dimensional
Allen-Cahn/Cahn-Hilliard system discussed in
section~\ref{sec:infin-dimens-appl} are of this type for their
deterministic dynamics, even though we simplified the noise terms in
both cases.

\subsection{Numerical aspects}

Conducting the numerical minimization procedure is non-trivial due to
the infinite time of the transition, $T=\infty$. We therefore use the
simplified geometric minimum action method
\cite{heymann-vanden-eijnden:2008,
  grafke-schaefer-vanden-eijnden:2016}, which harnesses the fact that
the minimal action is invariant under reparametrization, and performs
minimization in the space of arc-length parametrized curves
instead. See~\ref{sec:sgmam} for details on the procedure. This
computation is in contrast to the equilibrium setup, where the
heteroclinic orbits describe the transition completely, and are
efficiently computable by the string-method
\cite{e-ren-vanden-eijnden:2002,
  e-ren-vanden-eijnden:2007}. Interestingly, the string method can
still be used in the nonequilibrium context to identify the slow
manifold $\mathcal{M}$ of the slow-fast system at hand, as outlined
in~\ref{sec:slowmanifold}. It is important to note that our numerical
method relies neither on the large time scale separation nor on the
bifurcation structure of the system. Instead, it is applicable in
general for the computation of any large deviation minimizer. This
turns out to be of importance in applications, since this allows us to
handle arbitrarily complicated bifurcation structures and small but
finite values of $\alpha$, where the dynamics around bifurcation
points and close to the slow manifold might become very complicated
and might not be captured completely by the picture painted above. If
anything, large values of time scale separation, $\alpha\ll1$, are a
numerical annoyance rather than help because of their associated
stiffness.

\section{Generic examples}
\label{sec:generic-examples}

\subsection{Saddle-node bifurcation}

\begin{figure}[tb]
  \begin{center}
    \includegraphics[width=200pt]{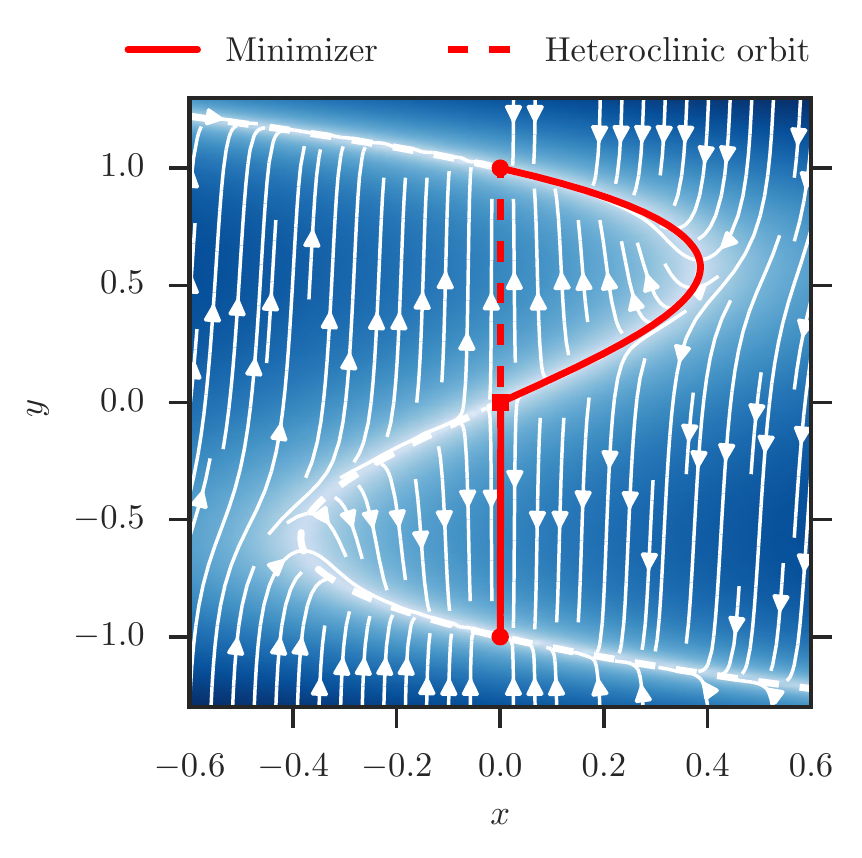}
    \includegraphics[width=225pt]{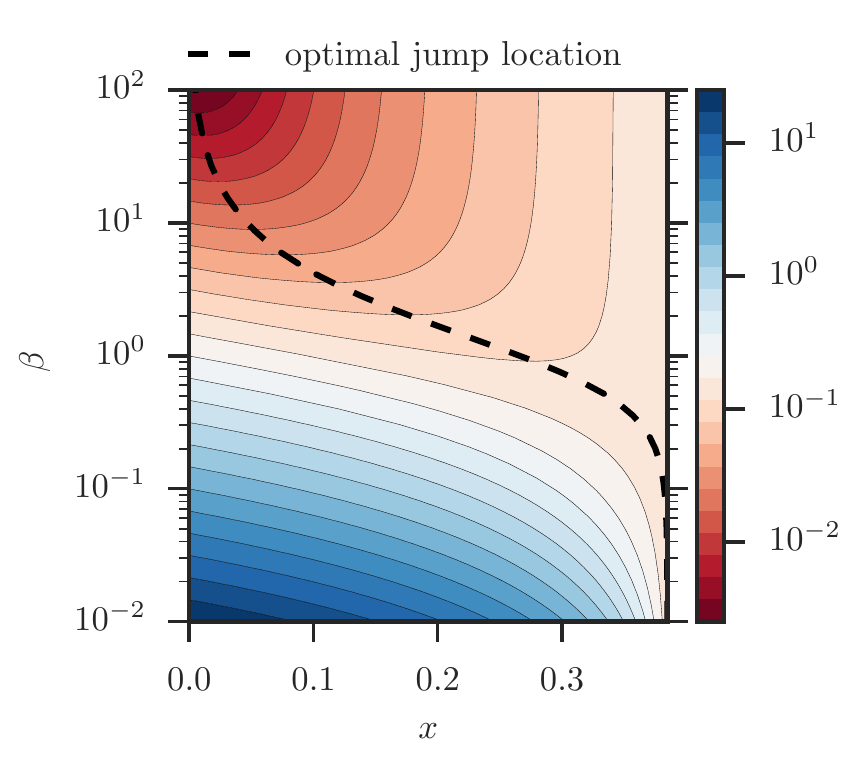}
  \end{center}
  \caption{Saddle-node bifurcation. \textit{Left:} Bifurcation diagram
    with the slow variable as control parameter. At the bifurcation
    point $z_X= {(-\tfrac29 \sqrt{3}, \tfrac13\sqrt{3})}$, a stable
    and an unstable branch of the slow manifold emerge. The MLP tracks
    the slow manifold into the saddle point. \textit{Right:} The
    optimal jump location as a function of the relative noise strength
    $\beta$ in the limit $\alpha\to0$.\label{fig:Stype-dynamics}}
\end{figure}

The prototypical example of a bifurcation is the saddle-node
bifurcation. In this section, we are going to construct a simple
metastable system out of two of these. Consider the stochastic
differential equation (SDE) for $z=(x,y)\in\RR^2$ given by
\begin{equation}
  \begin{aligned}
    d X &= -\alpha X dt + \sqrt{\alpha\epsilon}dW_x\\
    d Y &= \left(Y-Y^3-X\right)dt + \sqrt{\beta\epsilon}dW_y\,.
  \end{aligned}
\end{equation}
Note that this is the case with slow-fast noise, in which the noise
associated with the variables $X$ and $Y$ act on the slow and fast
timescales, respectively, as in~\eqref{eq:slowfastfgSDE}. Here we also
took $\sigma_x=1$ and $\sigma_y =\beta$, which therefore quantifies the
relative strength of the fluctuations on the slow and fast
variables. 

The deterministic system has two stable fixed points, which are
located at $z_A = (0,1)$ and $z_B = (0,-1)$ and a saddle point at
$z_S=(0,0)$. The $x$ component is invariant under the fast dynamics
and is therefore a slowly evolving quantity. It foliates the slow
manifold $\mathcal M$ and we can take it as the control parameter
$\mu$ for a bifurcation analysis. The slow manifold $\mathcal M$ is
comprised of all points $z_{\mathcal{M}}=(s-s^3, s)$ with $s\in\RR$,
since $f(z_{\mathcal M})=0$. For large negative $\mu$, $\mathcal{M}$
has a single stable branch. Increasing $\mu$, a saddle-node
bifurcation occurs at the point $z_X= {(-\tfrac29 \sqrt{3},
  \tfrac13\sqrt{3})}$, where another pair of branches emerge, one
stable and one unstable. The unstable branch then disappears for
positive $\mu$ in exactly the same way, leaving only a single stable
branch for large values of $\mu$.

This system violates detailed balance and is not of gradient type, and
therefore forward- and backward transitions between the stable fixed
points $z_A$ and $z_B$ will not occur along the same trajectory. Both
$z_A$ and $z_B$ lie on the slow manifold $\mathcal{M}$, and
$\mathcal{M}$ connects both fixed points. It therefore seems intuitive
that, in the small noise limit, $\epsilon\to0$, the transition
trajectory approaches the separatrix along the slow manifold, on which
the drift has amplitude $\mathcal O(\alpha)$, and is therefore easier
to overcome. This intuitive picture is confirmed in
Fig.~\ref{fig:Stype-dynamics} (left): Here, the deterministic dynamics
are depicted by the streamlines and their magnitude by the background
shading. The two stable fixed points are marked by points, and the
saddle by a square. The heteroclinic orbit connecting the saddle to
the fixed points necessarily is a straight line, represented by the
dashed line. The maximum likelihood transition pathway on the other
hand uses the slow manifold as a transition channel, and therefore
tracks $\mathcal{M}$ very closely from $z_A$ to $z_S$. It approaches
the separatrix between the two basins of attraction at the bifurcation
point $z_X$, and henceforth nearly deterministically tracks it to the
saddle. Only subsequently does it follow the deterministic dynamics
from the saddle onward.

It is important to keep in mind that depending on the physical
application the fluctuations associated with the reduced dynamics
might be of order $\mathcal O(\sqrt{\alpha})$ (as assumed in this
case). Then, even though the dynamics to overcome are weak in
comparison to the fast dynamics, so is the noise term itself. In the
limit $\alpha\to0$, the transition can then be considered as arising
from the combined effect of two distinct phenomena. The transition
either tracks the slow manifold on the slow time scale, using the
$\mathcal O(\sqrt{\alpha})$ fluctuations to overcome the $\mathcal
O(\alpha)$ dynamics, or it can jump from one branch of the slow
manifold to another, using the $O(1)$ fluctuations to overcome the
$\mathcal O(1)$ fast dynamics. In the associated action, both
contributions scale identically independently of $\alpha$, and the
actual path of the transition has to be identified by minimization of
this action. In general, the transition mechanism combines both
effects: A migration along $\mathcal M$ until an \emph{optimal jump
  location} is reached, and a subsequent jump from one branch to
another, reaching the separatrix to the other basin.

The optimal jump location can then be identified explicitly in terms
of $\beta$. Fig.~\ref{fig:Stype-dynamics} (right) shows the total
action $S_{\mathcal M} + S_{\text{jump}}$ as a function of the jump
location and $\beta$. The optimal jump location, where the minimum of
the total action is reached, will vary in $\beta$, with an almost
immediate jump at $\beta=10^2$, where the minimizer is basically
identical to the heteroclinic orbit, to essentially no jump at all at
$\beta=10^{-2}$. Fig.~\ref{fig:Stype-dynamics} (left) was obtained
using $\beta=10^{-1}$.

One should keep in mind that these considerations are only relevant if
the fluctuations of the reduced dynamics scale as $\mathcal
O(\sqrt{\alpha})$. In what follows, with the exception of the insect
outbreak model discussed in Sec.~\ref{sec:insect-outbreak}, a global
$O(1)$ noise is assumed. In these cases, no jump can happen as
$\alpha\to0$, and the transition will track the slow manifold
completely. In cases where part of the noise \textit{does} scale
$\mathcal O(\sqrt{\alpha})$, both the optimal jump location and the
actual transition behavior for finite $\alpha$ usually have to be
established using numerics, as done in this discussion.

The observed structure of the slow manifold $\mathcal M$ is
re-occurring in different physical applications, including all
applications discussed below: Two metastable fixed points are located
on two distinct stable branches, separated by the unstable branch of
$\mathcal M$, which coincides with the separatrix. Examples include the
FitzHugh-Nagumo model~\cite{fitzhugh:1961,
  nagumo-arimoto-yoshizawa:1962}, which describes the excitability of
the electrical potential across neural cell membranes in neural
dynamics, and the more complicated
Hodgin-Huxley~\cite{hodgkin-huxley:1952} model.

\subsection{Pitchfork bifurcation}
\label{sec:picthfork}

\begin{figure}[tb]
  \begin{center}
    \includegraphics[width=200pt]{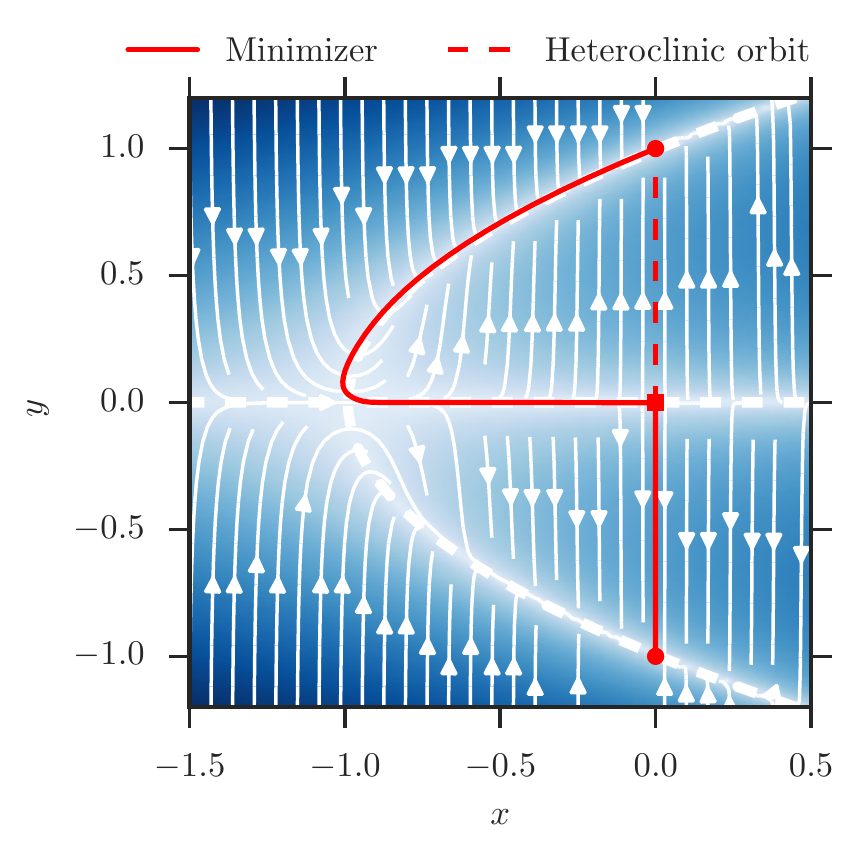}
    \includegraphics[width=200pt]{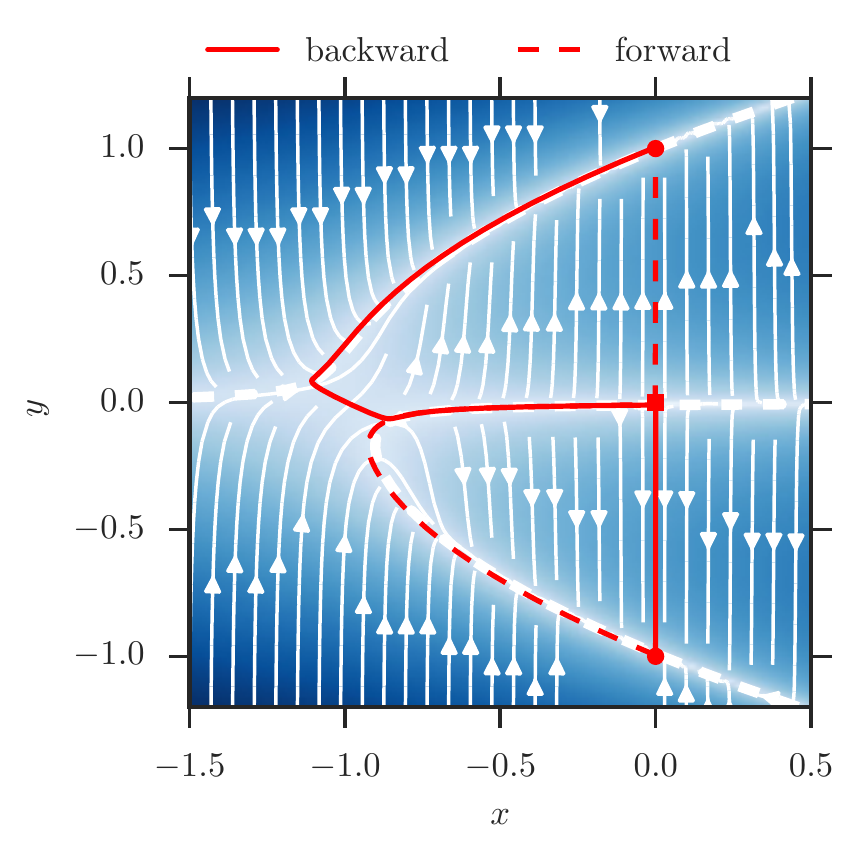}
  \end{center}
  \caption{Pitchfork bifurcation. \textit{Left:} The MLP tracks the
    slow manifold into the bifurcation point $z_X=(-1,0)$, while the
    heteroclinic orbit leaves the slow variable $x$
    unchanged. \textit{Right:} Introducing a tilt $z_\star$ separates
    the bifurcation structure, breaking the symmetry in the relative
    stability of the two fixed points. \label{fig:pitchfork}}
\end{figure}

A \emph{pitchfork bifurcation} is another example of a low-dimensional
bifurcation structure. Consider for example the SDE for $z=(x,y)\in\RR^2$
given by
\begin{equation}
  \label{eq:1}
  dZ = f(Z)\,dt + \alpha g(Z)\,dt + \sqrt{\epsilon} dW\,,
\end{equation}
with
\begin{equation}
  \label{eq:2}
    f(z) = 
    \left(\begin{matrix}
      0\\
      -y\left(y^2-x-1\right)
    \end{matrix}\right)\,\qquad
    g(z) = -z\,.
\end{equation}
Again the system violates detailed balance and is not of gradient
type. As before, $x$ is invariant under the fast dynamics and can be
taken as bifurcation parameter $\mu$ that foliates the slow manifold
$\mathcal M$. For large negative $\mu$, the system is only stable at
$y=0$. If we increase the bifurcation parameter, a supercritical
pitchfork bifurcation occurs at $\mu=-1$, where the single stable
branch of $\mathcal M$ splits into two stable and one unstable
branch. In~\eqref{eq:1} the noise is $\mathcal O(1)$ in $\alpha$, so
that the MLP, in contrast to the heteroclinic orbit, utilizes the slow
manifold for the transition. In particular, as depicted in
Fig.~\ref{fig:pitchfork} (left), it visits the bifurcation point
$z_X=(-1,0)$ and approaches the separatrix before reaching the saddle
point $z_S=(0,0)$.

The situation becomes interesting if we introduce a \emph{tilt}
$z_\star$ to the slow term, i.e.~choose $g(z)=z_\star - z$ instead
of~\eqref{eq:2}. Note that $g(z)$ modifies the drift of both $x$ and
$y$, even though it acts on the slow time scale $\mathcal O(\alpha)$
only. In the limit $\alpha\to0$, this tilt is therefore not felt, and
the slow manifold $\mathcal M$ remains identical to the situation
depicted in Fig.~\ref{fig:pitchfork} (left). Nevertheless, for any
finite choice of $\alpha$, this tilt will lead to a \emph{separation}
of the bifurcation structure and a breakdown of the pitchfork
bifurcation to a mere saddle-node bifurcation. In particular, only one
of the two separated components contains a bifurcation point and
consequently only one transition direction can make full use of the
slow manifold up to the saddle point, while the other has to bridge a
gap. The corresponding transition trajectories are depicted in
Fig.~\ref{fig:pitchfork} (right), with the backward trajectory (solid)
jumping from one part of the slow manifold to the other. In contrast,
the forward transition (dashed) can follow the slow manifold
completely to the bifurcation point and the saddle.

Even though the effect of the tilt is not felt as $\alpha\to0$, it
will have a dramatic effect for finite $\alpha$. % For the jump, the
% system has to overcome $\mathcal O(1)$ in $\alpha$ dynamics, to bridge
% across a gap of order $\mathcal O(\alpha |z_\star|)$. This results in
% an $\mathcal O(\exp(\alpha |z_\star|))$ change in the relative
% stability between $z_A$ and $z_B$.
For example, for the situation of Fig.~\ref{fig:pitchfork} (right),
$z_\star=(0,1)$ and $\alpha=10^{-2}$, but the ratio of the rates of
forward and backward transition between $z_A$ and $z_B$ changes by
roughly a factor $10^{5}$. An asymptotic analysis in $\alpha$ will be
oblivious to this effect, which highlights the necessity to perform a
numerical computation of the transition trajectory for practical
applications.

\section{Finite dimensional applications}
\label{sec:finite-dimens-appl}

\subsection{Insect outbreak}
\label{sec:insect-outbreak}

\begin{figure}[tb]
  \begin{center}
    \includegraphics[width=200pt]{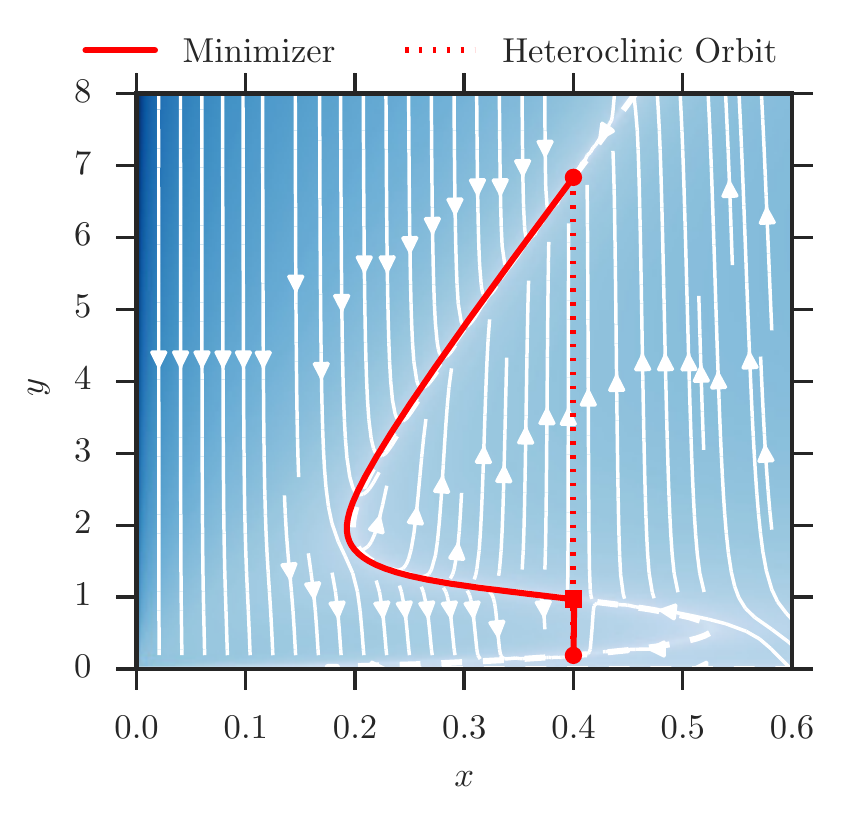}
    \includegraphics[width=200pt]{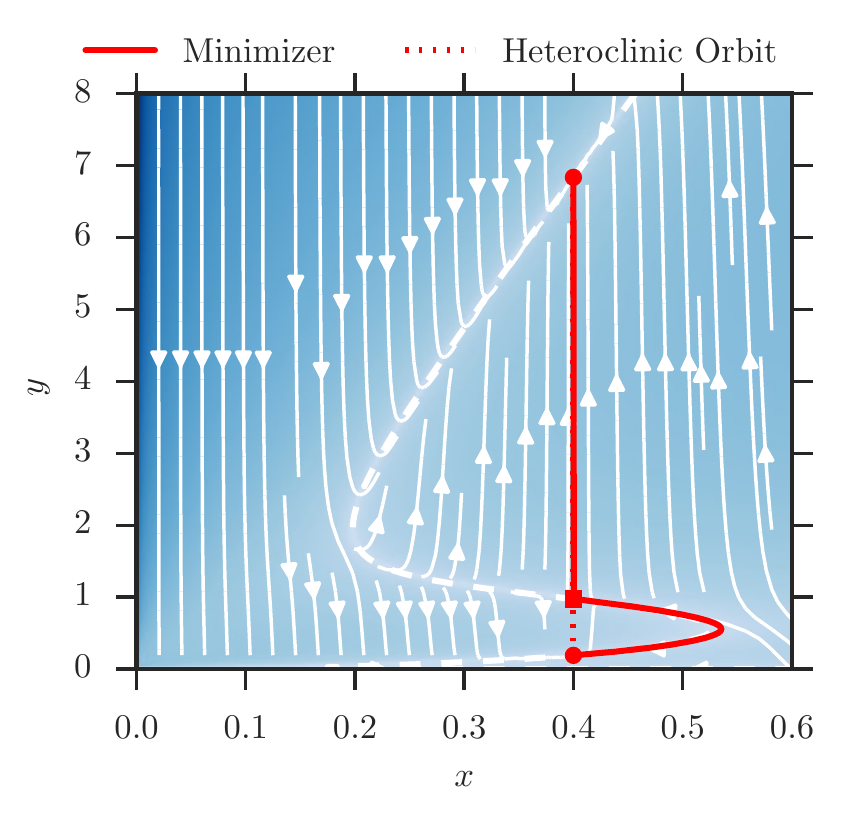}
  \end{center}
  \caption{Insect outbreak model, featuring a saddle-node bifurcation
    along the slow variable as control parameter. At the bifurcation
    point $z_X= (0.193, 1.96)$, a stable and an unstable branch of the
    slow manifold emerge. The MLP tracks the slow manifold through the
    bifurcation point into the saddle point. Depicted are forward
    (left) and backward (right) transitions in comparison to the time
    reversible heteroclinic orbit. \label{fig:insects}}
\end{figure}

A classical example of a slow-fast system from biology deals with
outbreaks of the spruce budworm which attacks the leaves of the balsam
fir tree. It was investigated extensively by
Ludwig~\cite{ludwig-jones-holling:1978, strogatz:2014} in the
deterministic context, who modeled the interaction between the slowly
recovering tree foliage area $X$ and the quickly reproducing budworm
population $Y$. In a slightly modified form, and adding stochastic
fluctuations to both degrees of freedom, we consider the system of
SDEs
\begin{equation}
  \begin{aligned}
    d X &= \alpha X(1-X/x_0)\,dt + \sqrt{\alpha\epsilon}\,dW_x\\
    d Y &= Y(1-\frac Y{Xy_0})\,dt - \frac{Y^2}{X^2+Y^2}\,dt + \sqrt{\epsilon}\,dW_y\,.
  \end{aligned}
\end{equation}
Here, the tree foliage area $X$ recovers slowly on time scale
$O(\alpha^{-1})$ to its long time limit $x_0$. In this simplified
model, the foliage is not influenced by the budworm's presence. For
the normalized budworm population $Y$, the carrying capacity of their
logistic growth, $X y_0$, depends on the tree foliage area
available. On top of that, budworms are subject to predation by birds,
which is modeled by the second term in the equation for $Y$. It is
chosen in a way that it saturates at high budworm densities due to
e.g.~territorial behavior of predators, but vanishes quadratically at
low densities to model learning and reward
mechanisms~\cite{ludwig-jones-holling:1978}. Both populations are
subject to stochastic fluctuations, which are assumed to be Gaussian
here for simplicity.

The SDE for $Z=(X,Y)\in\RR^2$ is readily treated by the fast-slow
formalism. The fast reproduction of the budworm (which can multiply
five-fold in a single year~\cite{strogatz:2014}) occurs on a time scale
of months, while the tree foliage area has a characteristic time scale
of decades. The budworm population near instantaneously reacts to a
change of the tree foliage area available; the resulting time scale
separation gives raise to a slow manifold $\mathcal M$ foliated by the
slow variable $x$, which is taken as control parameter $\mu$. In other
words, for a given available tree foliage area $x$, the budworm
population quickly adjusts to a compatible population density
$y_{\mathcal M}(x)$, which will correspond to a point on a stable
branch of $\mathcal M$ for this value of $x$. The tree foliage area
only slowly changes, until a metastable fixed point is
reached. Depending on the choice of parameters, the budworm/tree
system may exhibit multiple metastable fixed points. For example for
the choice $x_0=0.4$, $y_0=20$, $\alpha=10^{-2}$, two different
configurations of tree foliage area and budworm density each are
locally stable: $z_A = (0.4, 6.83)$ and $z_B=(0.4, 0.192)$, i.e.~one
configuration with a budworm outbreak, and one with a relatively low
presence of budworms. Furthermore, there is a saddle point at
$z_S=(0.4, 0.974)$. The slow manifold $\mathcal M$ is comprised of all
points $z_{\mathcal{M}}=(x, y_{\mathcal M}(x))$. For small values of
$\mu$, $\mathcal{M}$ has a single stable branch. Increasing $\mu$, a
saddle-node bifurcation occurs at the point $z_X= (0.193, 1.96)$,
where another pair of branches emerge, one stable and one
unstable. The unstable branch then disappears for large $\mu$ in
exactly the same way. The dynamics and the slow manifold $\mathcal M$
corresponding to this choice of parameters is depicted in
Fig.~\ref{fig:insects}, and are essentially those of the generic
saddle-node example above.

The stable fixed points $z_A, z_B$ lie on the slow manifold
$\mathcal{M}$, and $\mathcal{M}$ connects both fixed points. Here, the
maximum likelihood transition uses the slow manifold as a transition
channel, and therefore tracks $\mathcal{M}$ very closely from $z_A$ to
$z_S$. It approaches the separatrix between the two basins of
attraction at the bifurcation point $z_X$, and henceforth nearly
deterministically tracks it to the saddle. From the saddle onward, it
follows the deterministic dynamics. Depicted are the transitions in
both directions, $z_A\to z_B$ (left) and $z_B\to z_A$ (right). In both
cases the conclusion is that a transition is most probable through
slow fluctuations of the forest, instead of the insect population
undergoing unusually large population fluctuations to overcome the
barrier. This conclusion will no longer be true if either the
fluctuations of the insect population are much bigger than those of
the tree foliage area, or if the fluctuation strength becomes big
enough for the Kramers' time to be comparable to the slow time
scale. In both these situations, the most likely transition will
involve a fast portion, in which the path along the slow manifold is
shortcut by a jump between branches of this slow manifold, as
discussed in Sec.~\ref{sec:picthfork}.

\subsection{Phase separation with evaporation}
\label{sec:phase-separ-with}

\begin{figure}[tb]
  \begin{center}
    \includegraphics[width=200pt]{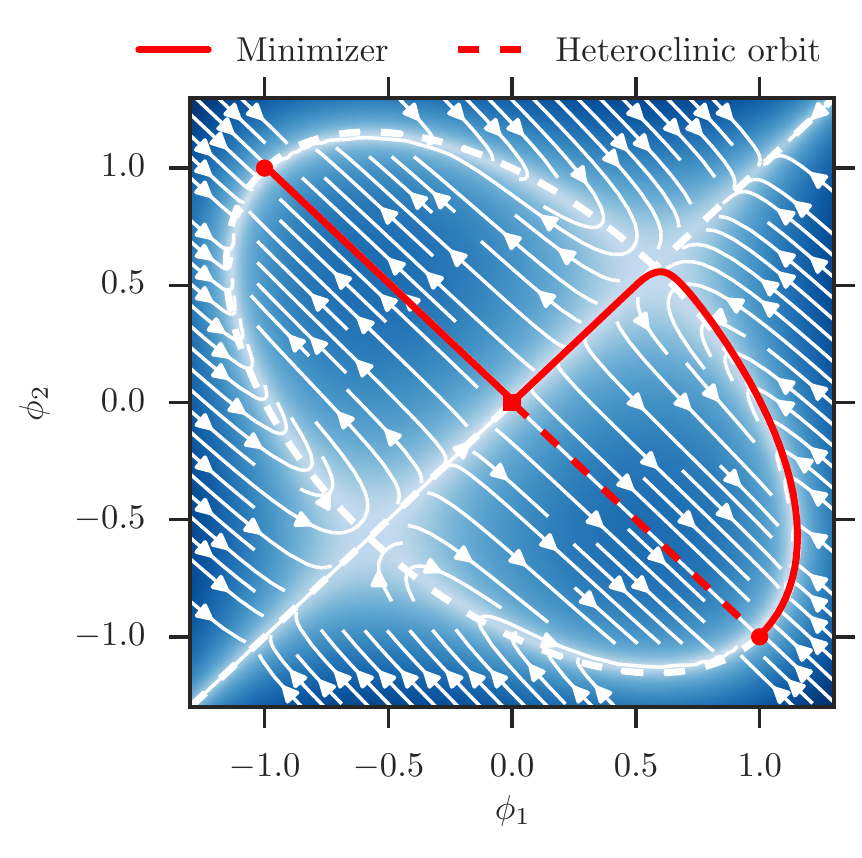}
    \includegraphics[width=200pt]{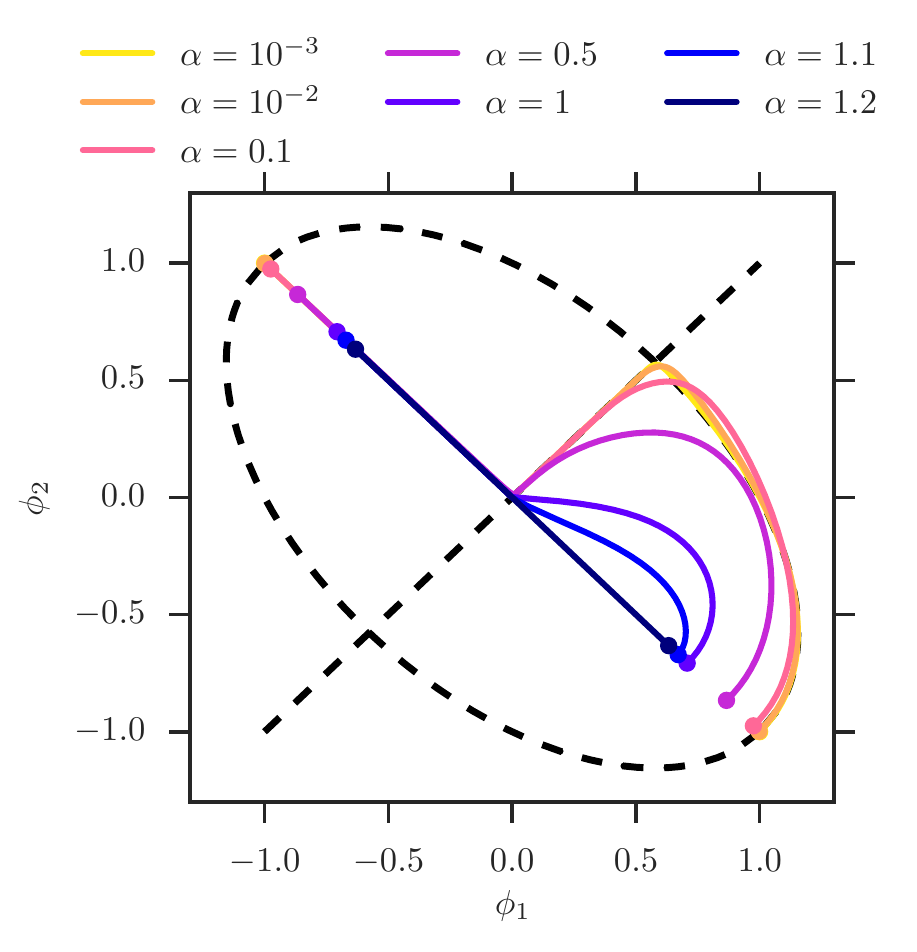}
  \end{center}
  \caption{Supercritical pitchfork bifurcation along the slow variable
    $\mu=\phi_1+\phi_2$. \textit{Left:} At the bifurcation point $\phi_X =
    (-\sqrt{2/3},-\sqrt{2/3})$, the stable slow manifold splits into
    two stable and one unstable branch. The MLP tracks the slow
    manifold through the bifurcation point $\phi_X$ into the saddle at
    $\phi_S=(0,0)$. \textit{Right:} Minimizers of the action
    functional \eqref{eq:action} for model~\eqref{eq:acch-toy} for
    different values of $\alpha$. For $\alpha\to0$, the transition
    happens increasingly close to $\mathcal
    M$. \label{fig:pitchfork-dynamics}}
\end{figure}
The next example describes a situation where the time scale separation
is introduced by two competing physical processes, namely \emph{phase
  separation} and \emph{evaporation}. We consider a simplified case
with only two degrees of freedom, $\phi_1$ and $\phi_2$, which
describe the concentration of some quantity relative to some reference
density in two neighboring compartments. Phase separation is modeled
by a free energy
\begin{equation}
  E_1(\phi) = \tfrac14 \sum_{i=1}^2 (1-\phi_i^2)^2\,,
\end{equation}
i.e.~a double well potential, which is combined with a conservative
mobility operator $M_1=Q=((1,-1),(-1,1))$, so that the mean
$\phi_1+\phi_2$ is conserved, and the system tends to the two minima
$(\pm 1,\mp1)$. This situation corresponds to most of the mass being
concentrated in either one of the two compartments, so that the
respective densities are above the reference density in one and below
the reference density in the other compartment. Evaporation is
considered slow in comparison, and is modeled by a reversal of both
degrees of freedom towards zero,
\begin{equation}
  E_2(\phi) = \tfrac12 \sum_{i=1}^2 \phi_i^2\,,
\end{equation}
with a mobility $M_2=\alpha\,\textit{Id}$. This leads to a reversal of
both densities towards the reference density, without necessarily
preserving the mean. The two terms compete on different time
scales. Due to the incompatibility of the mobility operators $M_1=Q$
and $M_2 = \alpha\, \mathit{Id}$, detailed balance is broken and
cannot be restored regardless of the choice of noise correlation. In
total, we can write the system as the SDE
\begin{equation}
  \label{eq:acch-toy}
  d\phi_i = \left(\sum_{j=1}^2 Q_{ij} (\phi_j-\phi_j^3) - \alpha\phi_i\right)dt + \sqrt{\epsilon} dW_i
\end{equation}
with $\phi=(\phi_1, \phi_2)$, and where we assumed a Gaussian noise on
all degrees of freedom, i.e.~dropping the noise associated with $E_1$,
$M_1$.

The dynamics of this model are depicted in
Fig.~\ref{fig:pitchfork-dynamics} (left): The streamlines describe the
direction of the deterministic dynamics, the shading its
magnitude. The two stable fixed points fulfill
$\lim_{\alpha\to0}\phi_A = (-1,1)$ and $\lim_{\alpha\to0} \phi_B =
(1,-1)$ and are separated by the separatrix $\phi_1=\phi_2$, the point
$\phi_S = (0,0)$ is a saddle point. In this case, the quantity $\mu =
\phi_1 + \phi_2$ remains unchanged under the fast dynamics. It is
therefore a slow quantity and is chosen to be our control
parameter. The slow manifold is comprised of all points where the fast
dynamics $Q(\phi-\phi^3)$ vanish. For large negative values of $\mu$,
this happens only on the straight line where $\phi_1=\phi_2$, which is
stable under the fast dynamics. The stability of the slow manifold
changes at $\mu=-\sqrt{1/3}$ and undergoes a supercritical pitchfork
bifurcation. The straight portion $\phi_1=\phi_2$ becomes unstable,
and two new stable branches form a tilted ellipse
$3(\phi_1+\phi_2)^2+(\phi_1-\phi_2)^2=4$. The same process occurs in
reverse for $\mu>0$, where stability of $\phi_1=\phi_2$ is reinstated
at $\mu=\sqrt{1/3}$. The complete slow manifold is depicted as a white
dashed line.
\begin{figure}[tb]
  \begin{center}
    \includegraphics[width=200pt]{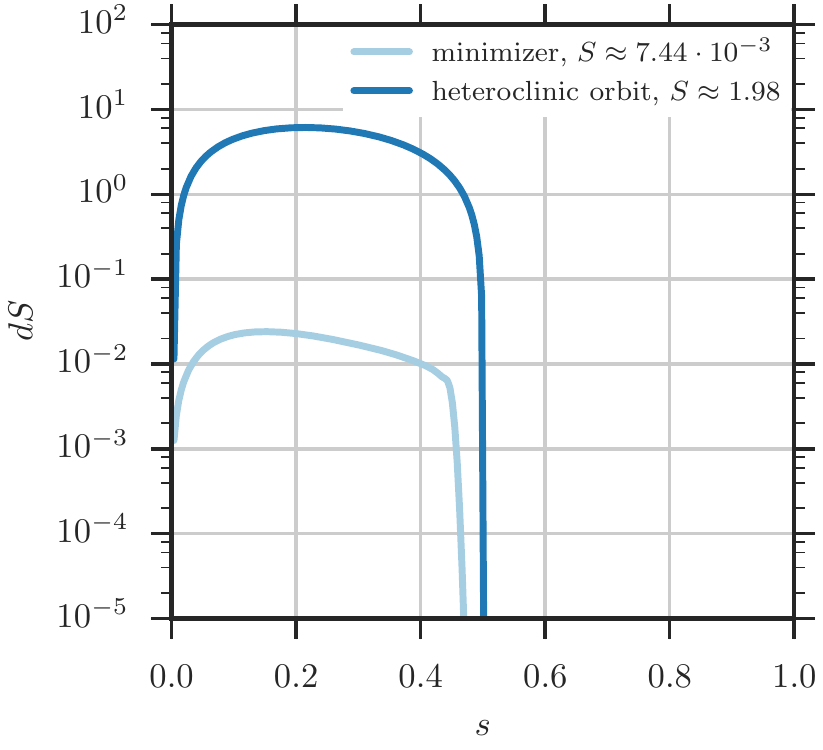}
    \includegraphics[width=200pt]{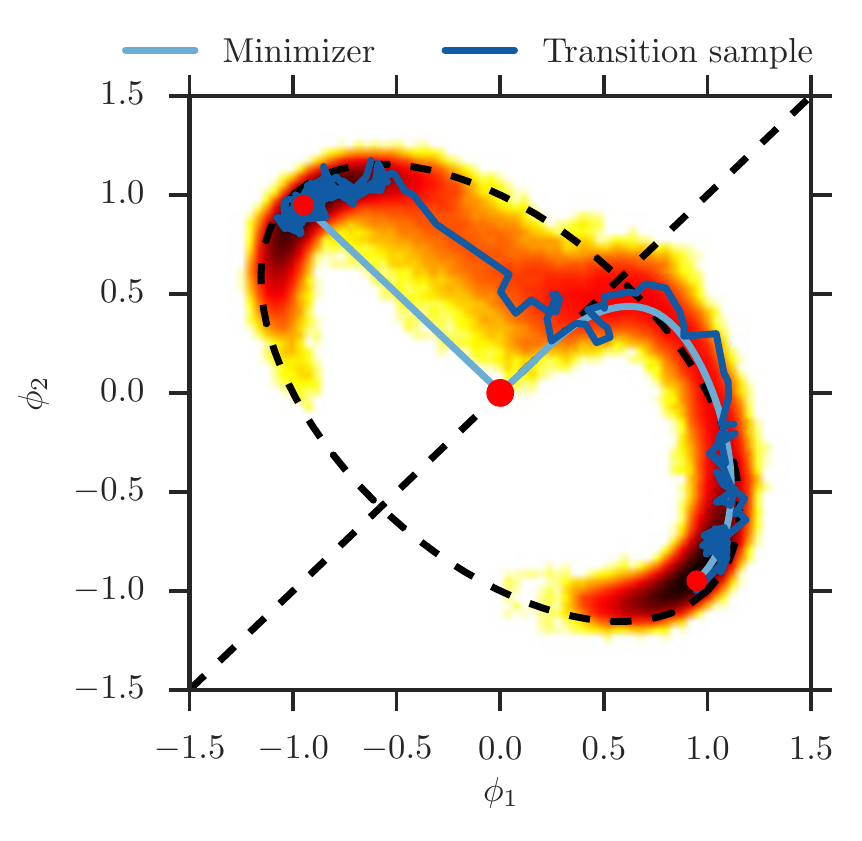}
  \end{center}
  \caption{\textit{Left:} Action density for the supercritical
    pitchfork bifurcation model. The path parameter is normalized to
    $s\in(0,1)$. The action is $\mathcal O(\alpha)$ smaller for the
    large deviation minimizer. \textit{Right:} Multiple transition
    trajectories obtained through direct sampling: The large deviation
    minimizer is followed up to the separatrix. The typical transition
    trajectory differs from the large deviation prediction from then
    on. Notably, almost no transition sample visits the transition
    state/saddle point in the center.\label{fig:acchtoy-action}}
\end{figure}
For model~\eqref{eq:acch-toy}, the minimizer is depicted by the red
solid line in Fig.~\ref{fig:pitchfork-dynamics}. As the minimizer
indeed follows the slow manifold, it approaches the separatrix at the
bifurcation point $\phi_X$, far from the saddle-point $\phi_S$. It
then tracks the separatrix quasi-deterministically into the
saddle-point to cross into the other basin of attraction and then
relax (deterministically) into the other fixed point. This is in
contrast to the heteroclinic orbit connecting the two fixed points,
which is the straight line through $\phi_A$ to $\phi_B$.

In the limit $\alpha\to0$ indeed we confirm numerically that the
transition trajectory for model \eqref{eq:acch-toy} approaches the
slow manifold. This fact is demonstrated in
Fig.~\ref{fig:pitchfork-dynamics} (right). Note also that the switch
to a straight line minimizer happens at a finite value $\alpha_c
\approx 1.12$, i.e. there is no continuous straightening of the
minimizer for growing $\alpha$, but a transition at a fixed critical
$\alpha=\alpha_c$. The action along the minimizer and the heteroclinic
orbit are depicted in Fig.~\ref{fig:acchtoy-action} (left). Notably,
due to its movement along the slow manifold, the action along the
minimizer is smaller by a factor $\mathcal{O}(\alpha)$ than the action
density along the heteroclinic orbit. This implies an exponentially
larger transition probability along $\mathcal M$.

Due to the presence of the slow manifold for $\alpha\ll1$, the
noise-induced transition will approach the separatrix between the
basins of attraction of the two stable fixed points not at the saddle
point, but on the bifurcation point. Note that even though in the
small noise limit $\epsilon \to 0$ the saddle point is visited
subsequently as well, this is no longer true for any finite noise. In
the presence of small but finite fluctuations, the typical transition
trajectory will take a rather different trajectory due to the fact
that the slow manifold becomes unstable after the bifurcation
point. Fig.~\ref{fig:acchtoy-action} (right) depicts this
effect. Shown are the large deviation minimizer and a transition
sample obtained by directly simulating the SDE~\eqref{eq:acch-toy}
with finite noise. The background shading indicates the probability
density for visiting each point conditioned on a transition
happening. The transition samples follow closely the minimizer and the
slow manifold up to the bifurcation point, but subsequently fall of
the separatrix before reaching the saddle point in the center. These
entropic effects imply that only the piece of the transition
trajectory obtained through large deviation theory that requires the
noise is observed in practice.  Still, both the transition
probabilities and the mean first passage times predicted by large
deviation arguments remain correct. As the transition behind the
bifurcation point is effectively deterministic, the minimum of the
large deviation rate function is not changed by these modified
transition trajectories.

\subsection{Tilted phase separation}

\begin{figure}[tb]
  \begin{center}
    \includegraphics[width=200pt]{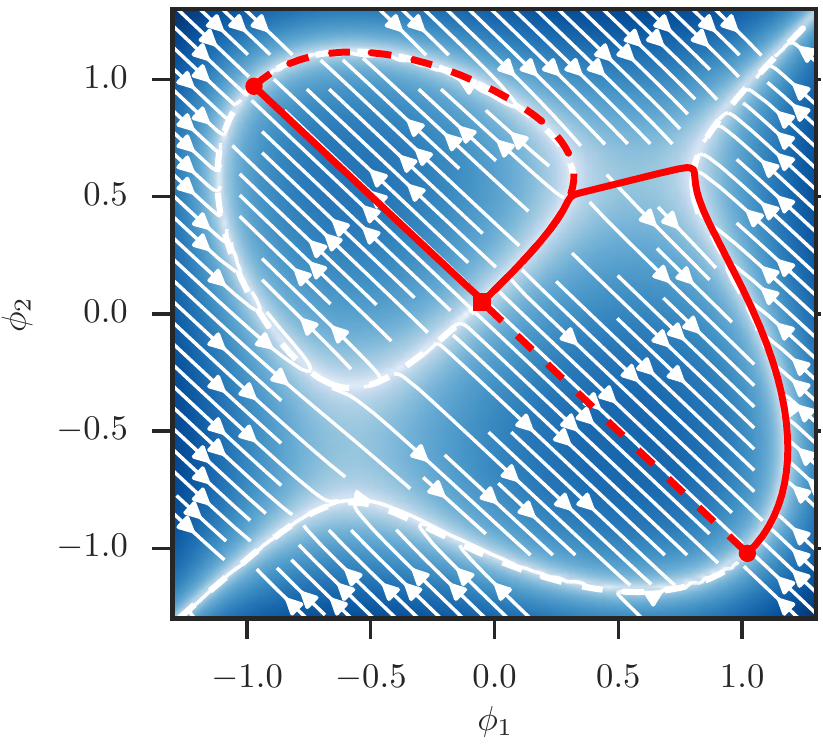}
    \includegraphics[width=200pt]{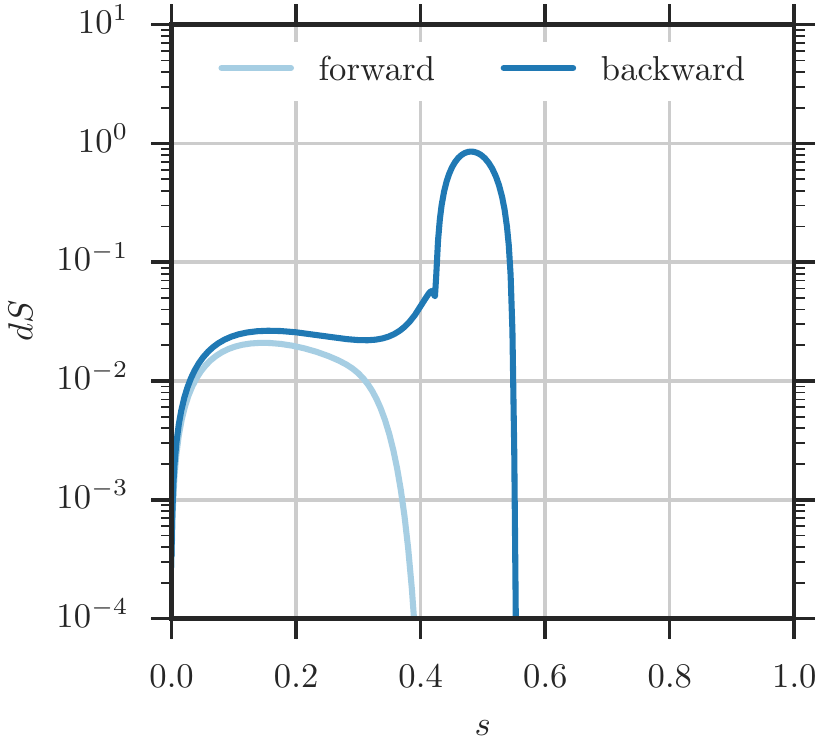}
  \end{center}
  \caption{Inhomogeneous tilt $\phi_\star=(0.1,-0.1)$. \textit{Left:}
    Forward and backward reaction differ, the slow manifold becomes
    disconnected. \textit{Right:} Action density for the corresponding
    forward and backward reaction, logarithmic in $dS$. The jump from
    one slow manifold to the other is clearly visible as peak in the
    backward action. The forward action is order $O(\alpha)$
    lower.\label{fig:tilted}}
\end{figure}

Consider a modification of \eqref{eq:acch-toy} where we tilt the
$E_2$ potential to break the symmetry between forward and backward
reaction,
\begin{equation}
  \label{eq:tilted}
  d\phi = \left(Q (\phi-\phi^3) - \alpha(\phi-\phi_\star)\right)dt + \sqrt{\epsilon} dW
\end{equation}
where $\phi_\star$ defines the tilt. An homogeneous tilt
(i.e.~$\phi_{\star,1} = \phi_{\star,2}$) does not change the relative
stability of the fixed points, since the system is still symmetric
under reflection at the separatrix $\phi_1 = \phi_2$. This is no
longer true if we tilt orthogonal to the separatrix, as depicted in
Fig.~\ref{fig:tilted} (left). Here, forward and backward transition
necessarily differ. At the same time, the breaking of the symmetry
implies a reduction of the bifurcation structure from a pitchfork
bifurcation to a mere saddle-node bifurcation. As a consequence, the
slow manifold becomes separated, and only one transition direction can
make full use of the slow manifold. Shown in Fig.~\ref{fig:tilted}
(right) is the action for the forward and backward reaction, with a
clear peak for the backward transition at the ``jump'' from one to the
other slow manifold. In contrast, the forward transition can follow
the slow manifold completely to the saddle, and its action is
therefore lower by a factor $10$. For a fixed finite $\alpha$ this
renders the probability to observe the system close to state
$\phi_B\approx(1,-1)$ exponentially higher. The system becomes trapped
on the isolated stable branch of $\mathcal M$ and will almost never
visit $\phi_A\approx(-1,1)$.

\section{Infinite dimensional applications}
\label{sec:infin-dimens-appl}

The examples shown so far where finite dimensional. In physics, system
of interest are often spatially extended, meaning that they have an
infinite number of degrees of freedom. In such a scenario, the
unknowns are given as functions on this space, and the finite
dimensional $\RR^d$ is replaced by a suitable function
space. Physically, the free energy is substituted by a \emph{free
  energy functional}, whose functional gradient, along with the
associated \emph{mobility operator} will make up the reversible
dynamics. The Euclidean norm is replaced by an appropriate norm on the
associated function space (e.g. the $L^2$ inner product and its
induced metric) and the fluctuation becomes spatio-temporal white
noise. Consequently, the obtained models will turn out to be
stochastic partial differential equations (SPDEs), instead of SDEs. It
is a non-trivial task to make mathematical sense of such SPDEs in
general: The possible ill-posedness of non-linear terms may require a
renormalization of the equation, which can (in some cases) be done
rigorously using the theory of regularity structures
\cite{hairer:2014}. In the context of LDT, one has to additionally
ensure whether the renormalization remains valid in the limit
$\epsilon\to0$. For more specific cases, the existence of an LDT can
be proved~\cite{faris-jona-lasinio:1982}. Here, we will not focus on
these aspects, and in the following assume that the functional
generalization of the action functional~\eqref{eq:action} is a valid
description of the transition behavior in the large deviation
sense. This assumption is generally taken to be true in cases of
practical interest, for example in macroscopic fluctuation theory
(MFT)~\cite{bertini-de_sole-gabrielli-etal:2015}.

On the numerical side, the infinite dimensional function space will be
truncated by discretization, which converges to the continuous
solution as the number of discretization points $N$ becomes large. As
a consequence, the minimization of the action functional has to be
undertaken in a vastly larger search space. It is in these examples in
particular, that the reduction to arc-length parametrized transition
trajectories is imperative. See~\ref{sec:sgmam} for details on the
implementation of such an optimization procedure.

\subsection{Allen-Cahn/Cahn-Hilliard dynamics}

\begin{figure}[tb]
  \begin{center}
    \includegraphics[width=290pt]{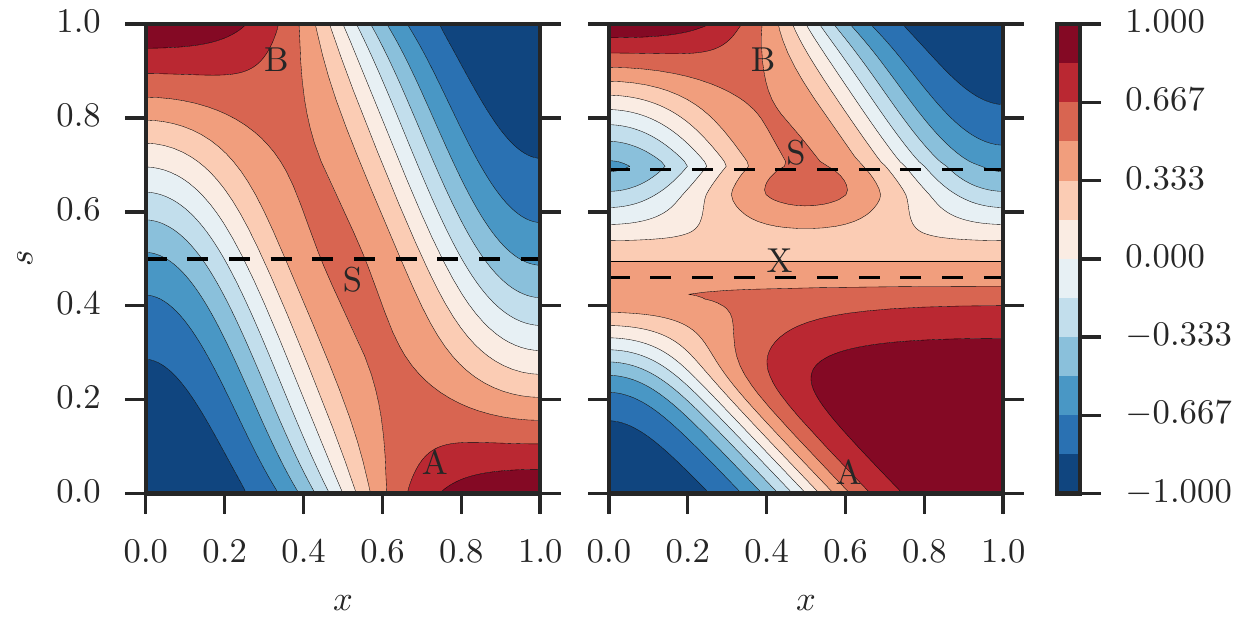}
    \includegraphics[width=150pt]{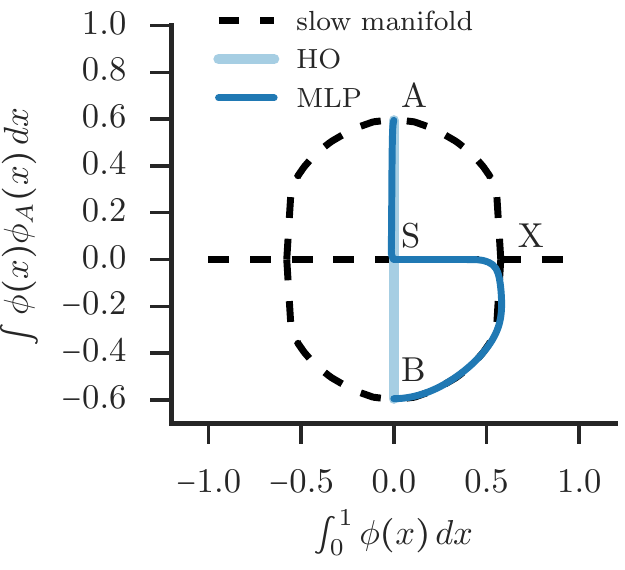}
  \end{center}
  \caption{Transition pathways between two stable fixed points of
    equation \eqref{eq:model-acch}. \textit{Left:} Heteroclinic orbit,
    defining the deterministic relaxation dynamics from the
    saddle-point $S$ down to $A$ and $B$. \textit{Center:} Maximum
    likelihood pathway, defining the most probable transition pathway
    from $A$ to $B$, following the slow manifold up to the bifurcation
    point $X$, and into the saddle $S$. \textit{Right:} Projection
    into a two-dimensional plane (see text). The stable fixed points
    are located at $A$ and $B$, the saddle point at $S$. The
    separatrix is the straight line through $S$ and $X$. The
    heteroclinic orbit (light blue) travels $A\to S\to B$ in a
    vertical line, while the minimizer (dark blue) travels first along
    the slow manifold (dashed) $A\to X$ into the bifurcation point $X$
    and then tracks the separatrix $X \to S$ into the
    saddle.\label{fig:transition}}
\end{figure}

\begin{figure}[tb]
  \begin{center}
    \includegraphics[width=200pt]{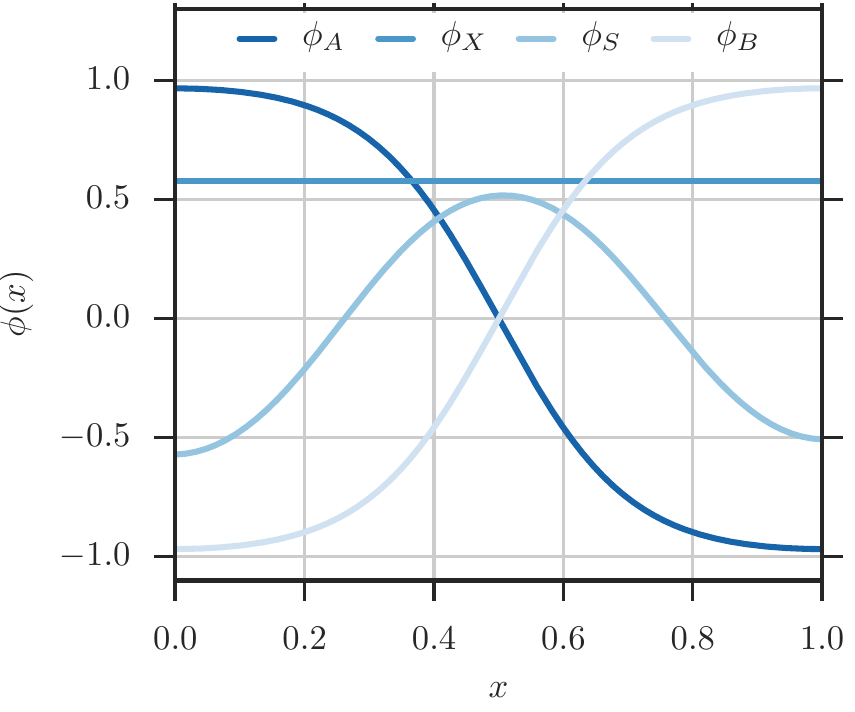}
    \includegraphics[width=200pt]{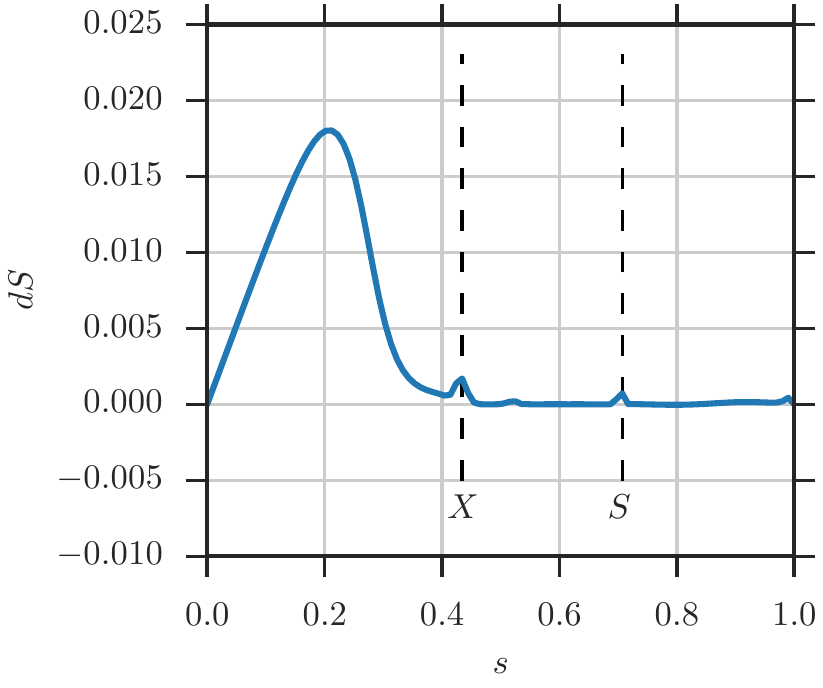}
  \end{center}
  \caption{\textit{Left:} The configurations $A, B, S, X$ in space:
    $\phi_A$ and $\phi_B$ are the two stable fixed points, $\phi_S$ is
    the saddle. At the bifurcation point $\phi_X$, the slow manifold
    intersects the separatrix. \textit{Right:} Action along the
    minimizer. Note that the action is non-zero climbing up the
    slow-manifold, but becomes zero already at the bifurcation point
    $X$, where it approaches the separatrix, before it reaches the
    saddle $S$.\label{fig:corners}}
\end{figure}

Consider the SPDE
\begin{equation}
  \label{eq:model-acch}
  \phi_t = \Proj(\kappa \phi_{xx} + \phi - \phi^3) - \alpha\phi
  + \sqrt{\epsilon}\,\eta(x,t), \qquad x\in [0,1]
\end{equation}
with Neumann boundary conditions $\phi_x(0)=\phi_x(1) = 0$ and where
$\kappa>0$, $\alpha>0$, and $\epsilon>0$ are parameters, $\eta(x,t)$
is spatio-temporal white-noise, and $\Proj$ is an operator with zero
spatial mean. For $\Proj = -\partial_{x}^2$ the system is a mixture of
a stochastic Allen-Cahn \cite{allen-cahn:1972} and Cahn-Hilliard
\cite{cahn-hilliard:1958} type dynamics. We will consider $\Proj(\phi)
= \phi - \int_0^1 \phi(x)\, dx$, which is similar in most discussed
aspects but simpler to investigate numerically. Note that for both
choices of $\Proj$, the fast dynamics are conservative, while the slow
dynamics are not. The slowly changing mean $\int_0^1 \phi(x)\,dx$ is
therefore taken as control parameter. Similar to the finite
dimensional case, we again represent all fluctuations by a single
Gaussian noise and do not consider the degenerate noise associated
with the conservative term.

This model is inspired by the scalar $\phi^4$ field theory for active
matter phase separation introduced by
\cite{wittkowski-tiribocchi-stenhammar-etal:2014}. In particular for
the ``Active Model B'' in
\cite{wittkowski-tiribocchi-stenhammar-etal:2014}, detailed balance of
a $\phi^4$ phase separation is mildly violated. A concrete application
might be the \textit{motility induced phase separation} (MIPS) of
actively propelled motile microorganisms~\cite{tailleur-cates:2008,
  cates-tailleur:2015}. Assuming a decreasing swim speed of motile
bacteria such as \textit{E.~coli} with increasing local density of the
bacteria, a feedback loop is created. Accumulation of bacteria leads
to their slow down, which in turn induces further accumulation. The
resulting phase separation can be combined with reproduction to obtain
rich phenomenology and spatio-temporal patterns reminiscent of the
biofilm-planktonic lifecycle observed in
nature~\cite{cates-marenduzzo-pagonabarraga-etal:2010,
  grafke-cates-vanden-eijnden:2017}. Similar in spirit, our model
consists of a phase-separating $\phi^4$ free energy functional and a
linear restoring term modeling convergence to a carrying capacity. The
corresponding free energies are given by
\begin{subequations}
  \label{eq:form-acch}
  \begin{align}
    E_1(\phi)&=\int_0^1 \left(\frac12 \kappa |\phi_x|^2 
               - \frac12 |\phi|^2 + \frac14
    |\phi|^4\right) dx,& M_1&=P\\
    E_2(\phi) &= \frac12 \int_0^1  |\phi|^2 dx,& M_2&=\alpha\,\textit{Id}\,.
  \end{align}
\end{subequations}
The deterministic dynamics ($\epsilon=0$) involves a competition
between the drift associated with $E_1$, which tends to separate the
field toward $\phi_\pm=\pm1$ in a conservative way, and that
associated with $E_2$, which tends to bring it towards $\phi=0$ in a
non-conservative way.  We will see below that the net effect of this
competition is that the deterministic system admits no constant stable
solution if $\alpha$ is small enough.

As argued before, the degeneracy of the conservative mobility yields a
slow control parameter for a bifurcation analysis. In particular, at
finite $\epsilon$ (i.e. with the effect of the noise included), the
competing dynamics violate detailed balance and the mechanisms of
forward- and backward-transitions between metastable states
differ. These metastable states are the solutions of
\begin{equation}
  \label{eq:fixedp}
   \Proj(\kappa\phi_{xx}+\phi-\phi^3)-\alpha \phi = 0\,.
\end{equation}
The only constant solution of this equation is the trivial fixed point
$\phi(x)=0$, whose stability depends on $\alpha$ and $\kappa$. In the
following, we choose $\alpha=10^{-2}$ and $\kappa = 2\cdot 10^{-2}$,
and therefore are in the regime where $\phi(x)=0$ is unstable. Two
stable fixed points obtained by solving \eqref{eq:fixedp} for these
values of $\alpha$ and $\kappa$ are depicted in Fig.~\ref{fig:corners}
(left) as $\phi_A$ and $\phi_B$, with $\phi_A = -\phi_B$.

The slow manifold $\mathcal{M}$ for this model is made of the
solutions of
\begin{equation}
  \label{eq:slowmanif-acch}
  \Proj (\kappa \phi_{xx}+\phi-\phi^3)=0\,.
\end{equation}
On this manifold the motion is driven solely via the slow term,
$-\alpha \phi$, on a time scale of order $O(\alpha)$, and the
noise. Equation~\eqref{eq:slowmanif-acch} can be written as
\begin{equation}
  \kappa \phi_{xx} + \phi - \phi^3 = \lambda
\end{equation}
where $\lambda$ is a parameter. As a result the slow manifold can be
described as one-parameter families of solutions parametrized by
$\lambda\in\mathbb{R}$ -- in general there is more than one family
because the manifold can have different branches corresponding to
solutions of~\eqref{eq:fixedp} with a different number of domain
walls.

Indeed, the two-dimensional pitchfork bifurcation SDE model
\eqref{eq:acch-toy} can be seen as a two-dimensional approximation of a
discretized version of \eqref{eq:model-acch}: The projection $\Proj
v=v-\int_0^1 v\,dx$ reduces to $\frac12 Q$ for a discretization
with $N_x=2$ grid points. The projection $Pv = -\partial_x^2 v$ for
periodic boundary conditions reduces to $2Q$ for the standard finite
difference 3-point Laplace stencil for $N_x=2$. In that sense, the
two-dimensional model \eqref{eq:acch-toy} is a discrete approximation
of \eqref{eq:model-acch} for either choice of $P$ with only $N_x=2$
degrees of freedom (when furthermore dropping the dissipation term,
$\kappa=0$).

Fig.~\ref{fig:transition} (left) shows the heteroclinic orbit
connecting the two stable fixed points: Along the complete trajectory,
the mean is preserved. The transition follows a domain wall motion in
the center, with a nucleation event at the boundary. The saddle point
$\phi_s$, denoted by $S$, which also demarcates the position at which
the separatrix is crossed, is the spatially symmetric configuration
with a central region at $\phi=\phi_+$ and two regions at the boundary
approaching $\phi=\phi_-$. In contrast, Fig.~\ref{fig:transition}
(center) shows the minimizer of the action functional
\eqref{eq:action}, representing the most probable transition path for
$\epsilon\to0$. Starting again at the fixed point $A$ this minimizer
takes a different course, moving the domain wall, at vanishing cost
for $\alpha\to0$, without inducing nucleation. At the point $X$, close
to the bifurcation point, the motion changes, tracking closely the
separatrix (which is identical to the unstable portion of the slow
manifold) into the saddle $S$. Note that the motion $X\to S$ happens
quasi-deterministically despite introducing two nuclei at the
boundaries, because of the slow drift pulling the mean towards $0$,
while the projected term is $0$ exactly.  From the saddle onward, $S
\to B$, the transition necessarily follows the heteroclinic orbit,
which is equivalent to the deterministic relaxation path. In this
respect, the SPDE model \eqref{eq:model-acch} exactly corresponds to
the two-dimensional model \eqref{eq:acch-toy}.

To further illustrate the resemblance to the two-dimensional model, we
choose to project the minimizer, the heteroclinic orbit and the slow
manifold onto two coordinates,
\begin{enumerate}[(i)]
\item the mean $\int_0^1 \phi(x) \,dx$, which corresponds to the
  direction $\phi_1+\phi_2$ of the two-dimensional model, and
\item the component of $\phi(x)$ in the direction of $\phi_A$ given by
  $\int_0^1 \phi(x)\phi_A(x)\,dx$, which corresponds to the direction
  $\phi_1-\phi_2$ of the two-dimensional model.
\end{enumerate}
The comparison of the transitions in the reduced coordinates is
depicted in Fig.~\ref{fig:transition} (right). This figure is not a
schematic, but the actual projection of the heteroclinic orbit and the
minimizer of Fig.~\ref{fig:transition} (left, center) according to (i)
and (ii) above. In this projected view, the separatrix is the straight
vertical line, and the pitchfork bifurcation structure of the slow
manifold is clearly visible, $X$ denoting the bifurcation point. The
movement of the minimizer (dark blue) along the slow manifold (dashed)
$A\to X$ and along the separatrix $X\to S$ (which coincides with the
unstable part of the slow manifold) into the saddle $S$ highlights its
difference to the relaxation pathway (light blue). The configurations
at the points $A, B, S$ and $X$ are depicted in Fig.~\ref{fig:corners}
(left).

To demonstrate that the motion along the minimizer becomes
quasi-deterministic already at $X$ before it hits the saddle at $S$,
Fig.~\ref{fig:corners} (right) shows the action density $dS$ along the
transition path. This quantity becomes nearly zero already at $X$.

\subsection{Tilted Allen-Cahn/Cahn-Hilliard model}

\begin{figure*}[tb]
  \begin{center}
    \includegraphics[width=290pt]{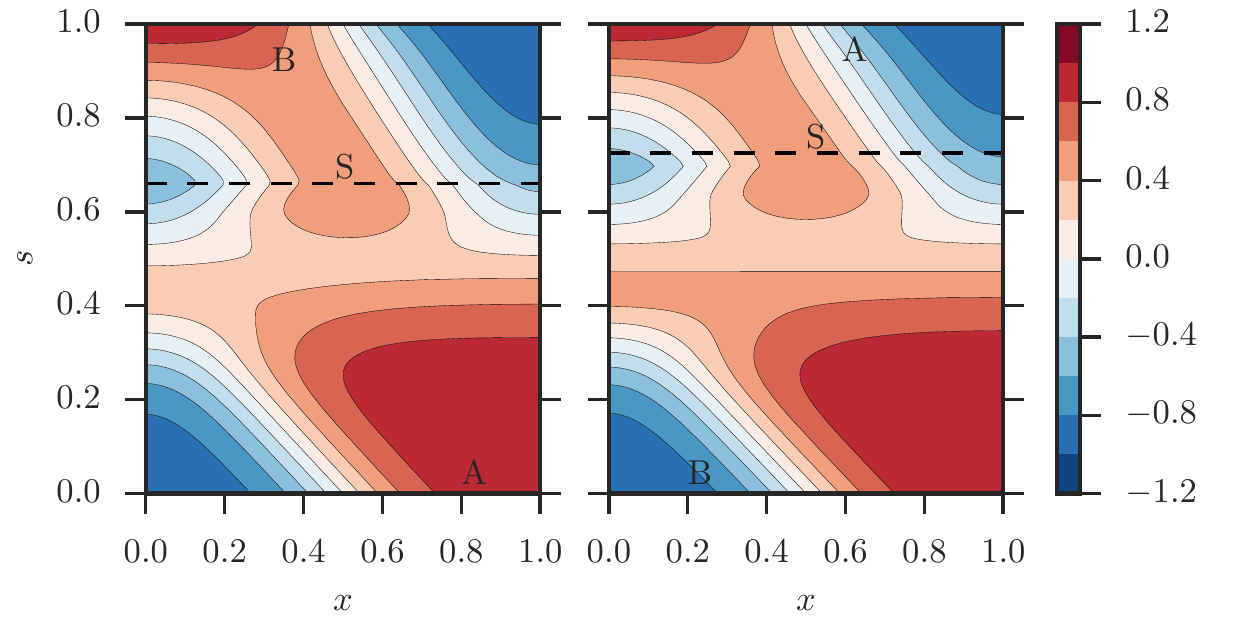}
    \includegraphics[width=150pt]{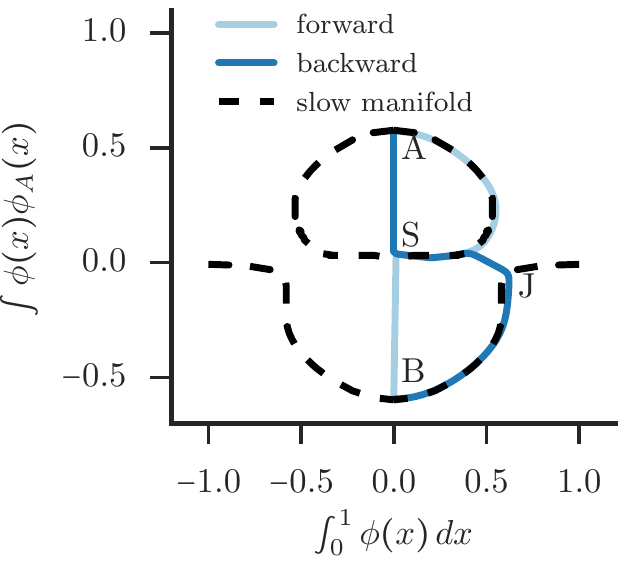}
  \end{center}
  \caption{Forward \textit{(left)} and backward \textit{(center)}
    transitions in the tilted PDE model, $\phi_\star=3\cdot
    10^{-2}\cos(x)$. \textit{Right:} Projection of the forward and
    backward reactions into the bifurcation diagram. Note that the
    slow manifold becomes separated under the tilt and the forward
    transition becomes exponentially more likely than the backward
    transition. \label{fig:pdetilt-cont}}
\end{figure*}

\begin{figure}[tb]
  \begin{center}
    \includegraphics[width=200pt]{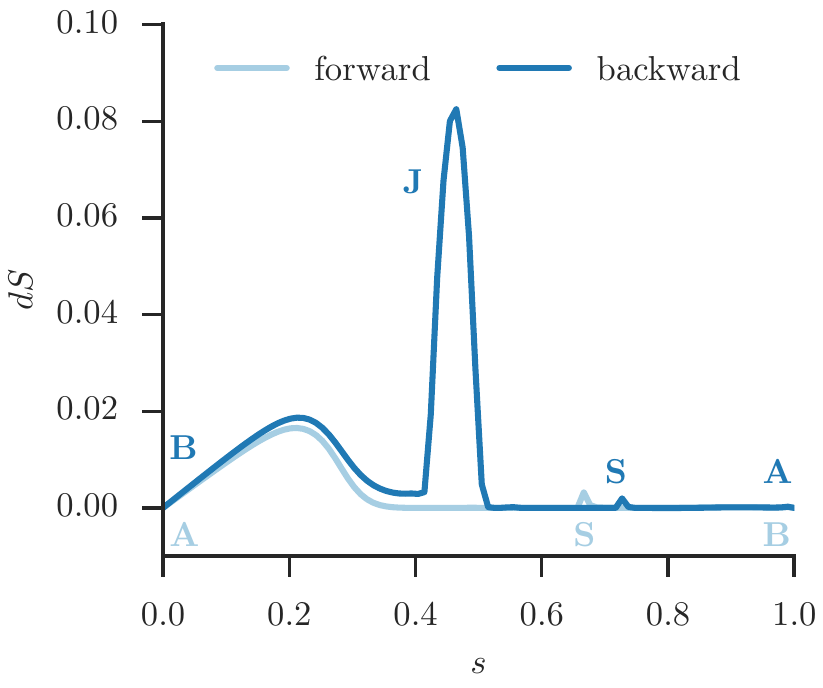}
  \end{center}
  \caption{Action density along forward and backward reaction. The
    jump at ``J'' between the slow manifolds leads to an $\mathcal O
    (\alpha)$ smaller action for the forward
    transition.\label{fig:pdetilt-action}}
\end{figure}

Similar to the two-dimensional reduction, the SPDE
model~\eqref{eq:model-acch} can also be tilted. Consequently, we are
taking
\begin{equation}
  \label{eq:model-tilt}
  \phi_t = \Proj(\kappa \phi_{xx} + \phi - \phi^3) - \alpha(\phi -
  \phi_\star) + \sqrt{\epsilon}\,\eta(x,t), \qquad x\in [0,1]\,.
\end{equation}
Again, a spatially homogeneous tilt $\phi_\star = \textrm{const.}$
will not result in a change in relative stability of the fixed points
since the transitions and fixed points are still symmetric under
$x\leftrightarrow1-x$. If instead we tilt in a spatially inhomogeneous
manner, we reproduce the segregation of the slow manifolds into two
disconnected components, and the relative stability of the fixed
points will change. The forward and backward transitions for the tilt
$\phi_\star=3\cdot 10^{-2}\cos(\pi x)$ is depicted in
Fig.~\ref{fig:pdetilt-cont} as well as its projection into the
bifurcation diagram depicting the slow manifold of the system. The
forward transition, starting from $A$, completely follows the slow
manifold into the saddle $S$, to then move into the other fixed point
$B$. In the opposite direction, before hitting the separatrix, the
trajectory has to jump from one branch of the slow manifold to another
at $J$. This jump is visible as a peak in the action density in
Fig.~\ref{fig:pdetilt-action}. Note that the total action for forward
and backward transition are drastically different, resulting again in
an exponential difference in the relative stability of the two
metastable fixed points. This is despite the fact that the spatial
variation of the tilt is very small.

The interpretation in the context of motile bacteria and MIPS is the
following: The spatially varying tilt $\phi_\star$ corresponds to a
small spatial perturbation in the carrying capacity, i.e. to small
spatial inhomogeneities in the availability of resources to sustain
bacteria. Even tiny spatial variances lead to a separation of the slow
manifold into multiple disconnected components. Whenever such a
disconnected component does not contain a bifurcation point, or
equivalently does not touch the separatrix, it will be easy to find by
the noisy exploration of the dynamics, but hard to leave again. This
translates into an exponentially amplified capability of motile
bacteria to locally cluster around slightly increased food sources.

\section{Conclusions}
\label{sec:conclusion}

From a conceptual viewpoint at least, noise-induced transitions are
relatively easy to analyze in systems whose dynamics satisfy
detailed-balance (microscopic irreversibility). Indeed, large
deviation theory indicates that, when the noise amplitude is small
compared to the height of the barriers of the potential over which
these systems navigate, the transitions proceed with high probability
via reaction channels made out of heteroclinic orbits that connect
local minima of the potential via saddle points. The situation is much
more complex in systems whose dynamics are not in detailed-balance.
In this set-up, there is no underlying potential, and the
noise-induced transitions are intrinsically out-of-equilibrium
phenomena. While large deviation theory still indicates that these
transitions proceed by preferred channels (namely, the minimizer of
the Freidlin-Wentzell action functional), these channels are not
heteroclinic orbits in general, and their identification is nontrivial
because of the wide variety of pathways they may take. While a general
theory of such transitions remains elusive, here we have shown that
some of their generic features can be identified in specific
situations: namely slow-fast systems whose slow manifold displays
bifurcations. In these systems, the slow manifold creates a preferred
channel of reaction between the metastable states that typically
differs vastly from the equilibrium prediction given by the
heteroclinic orbit. The minimizer of the Freidlin-Wentzell action uses
this channel, leading to an action that is lower by a factor
$\alpha^{-1}$ compared to the one calculated along the heteroclinic
orbit. Depending on the structure of the noise, this channel may or
may not be used all the way through, but its existence allows one to
develop robust numerical tools for the calculation of the actual
minimizer of the Freidlin-Wentzell action.  Such tools were proposed
here. Interestingly, they apply not only in the limit of infinite time
scale-separation, $\alpha \to 0$, but they can also be used in
situations where a small but finite $\alpha$ may create order one
discrepancies in the slow manifold structure that affect both the path
of transition and the action along it. We expect these numerical tools
to be therefore applicable in a wide range of situations with strong
but not infinite time scale separation between slow and fast
variables.

\section*{Acknowledgment}

EVE is supported in part by the Materials Research Science and
Engineering Center (MRSEC) program of the National Science Foundation
(NSF) under award number DMR-1420073 and by NSF under award number
DMS-1522767.

\appendix

\section{Numerical computation of the transition trajectory}
\label{sec:sgmam}

Consider the SDE
\begin{equation}
  dZ = b(Z)\,dt + \sqrt{\epsilon}\sigma(Z)\,dW\,,
\end{equation}
for $Z\in\RR^n$, which includes the slow-fast
dynamics~\eqref{eq:slowfast} as a special case. Here, $W$ is a Wiener
process on $\RR^n$, $b:\RR^n\to\RR^n$ describes the deterministic
dynamics and $\sigma: \RR^n\to \RR^n\times\RR^n$ denotes the noise
correlation. Large deviation theory specifies that the probability for
the process $Z$ to end up in a set $B\subset \RR^n$ is given by
\begin{equation}
  \label{eq:transprob}
  \PP^{z_0} (Z(T)\in B) \asymp \exp \big( -\epsilon^{-1} \min_{\psin\in\mathcal C} S_T(\psin)\big)\,,
\end{equation}
where $\asymp$ denotes log-asymptotic equivalence and the minimum is
taken over the set of functions $\mathcal C = \{\psin \in
C([0,T],\RR^n) : \psin(0)=z_0, \psin(T)\in B\}$. The action functional
$S_T(\psin)$ is given by
\begin{equation}
  S_T(\psin) = \int_0^T L(\psin,\dot\psin)\,dt
\end{equation}
for the Lagrangian
\begin{equation}
  \label{eq:Lag}
  L(\psin,\dot\psin) = \tfrac12|\sigma^{-1} (\dot \psin - b(\psin))|^2\,.
\end{equation}
LDT furthermore predicts the most likely transition pathway (MLP), or
more precisely
\begin{equation}
  \lim_{\epsilon\to0} \PP^x(\sup_{t\in[0,T]} |Z(t)-\psin_\star(t)| < \delta\ |\ Z(T)\in B)=1, \quad \forall\,\delta>0\,.
\end{equation}
The MLP is given by the minimizer of the action functional,
\begin{equation}
  \label{eq:optim}
  \psin_\star = \mathop{\text{argmin }}_{\psin\in\mathcal C} S_T(\psin)\,.
\end{equation}
From a numerical viewpoint, finding the transition
probability~\eqref{eq:transprob} and the most likely transition
$\psin_\star$ itself reduces to the \emph{deterministic} optimization
problem given by~\eqref{eq:optim}. Since the search space is a
function space, this optimization problem can become quite large, in
particular if the underlying system is spatially continuous (the SPDE
case). In principle, though, it can be solved by numerical
optimization techniques.

The optimization problem~\eqref{eq:optim} becomes particularly
intricate if we furthermore do not want to prescribe a transition time
$T$, but instead are interested for example in the relative
probability of the system to be found close to a given metastable
fixed point or the mean first passage time between different
metastable fixed points. These quantities are precisely defined in the
context of LDT through the \emph{quasipotential}
\begin{equation}
  \label{eq:QP}
  V(z_A,z_B) = \inf_{T>0} \min_{\psin\in\mathcal C_{z_A,z_B}} S_T(\psin)\,,
\end{equation}
where $\mathcal C_{z_A,z_B} = \{\psin\in C([0,T],\RR^n):\psin(0)=z_A,
\psin(T)=z_B\}$. Then, the mean first passage time $\tau_{A\to
  B}=\inf\{t>0 : Z(t)\in B_\delta(z_B), Z(0)=z_A\}$ between $z_A$ and
a ball of radius $\delta\ll1$ around $z_B$ is given by
\begin{equation}
  \EE \tau_{A\to B} \asymp \exp(\epsilon^{-1} V(z_A,z_B))\,,
\end{equation}
and the relative probability to find the system close to $z_A$ and
$z_B$ by
\begin{equation}
  \frac{p_A}{p_B} \asymp \exp(\epsilon^{-1}(V(z_A,z_B)-V(z_B,z_A))\,.
\end{equation}
Unfortunately, in general, the minimization over $T$ in the
computation of the quasipotential~\eqref{eq:QP} will not be
attained,~i.e.~$T\to\infty$. Numerically, this implies that we need to
discretize an infinitely long time interval, which complicates the
computation considerably.

An effective solution to this difficulty was proposed
in~\cite{heymann-vanden-eijnden:2008,
  grafke-schaefer-vanden-eijnden:2016} by considering a geometric
reformulation of the problem. Effectively, the computation of the
quasipotential can also be expressed as
\begin{equation}
  \label{eq:QP2}
  V(z_A,z_B) = \min_{\psin \in \mathcal C_{z_A,z_B}} \sup_{\theta: H(\psin,\theta)=0} \int_0^1 \langle \dot \psin,\theta\rangle \,ds\,,
\end{equation}
where $H(\psin,\theta)$ is the \emph{Hamiltonian} corresponding to the
Lagrangian~\eqref{eq:Lag} defined through
\begin{equation}
  H(\psin,\theta) = \langle b(\psin),\theta\rangle + \tfrac12 |\sigma(\psin) \theta|^2\,,
\end{equation}
and $\langle\cdot,\cdot\rangle$ denotes an appropriate inner
product. In essence, the formulation~\eqref{eq:QP2} can be understood
as Maupertius' principle from classical mechanics: The integral
in~\eqref{eq:QP2} no longer integrates over a (possibly infinite) time
interval, but instead is \emph{independent of the parametrization} of
the trajectory $\psin$. As a consequence, the Euler-Lagrange equations
are replaced by a least action principle over \emph{arc-length
  parametrized} trajectories. The numerical difficulty of an
optimization problem on infinite time horizons is reduced to a
minimization problem on geodesics of finite length.

The numerical computation of the large deviation minimizers for all
applications in this paper make use of this approach. For the
associated optimization problem, we use the algorithm presented
in~\cite{grafke-schaefer-vanden-eijnden:2016}, which generalizes those
introduced in~\cite{ren2004minimum,heymann-vanden-eijnden:2008}.

\section{Numerical computation of the slow manifold}
\label{sec:slowmanifold}

The slow manifold
\begin{equation}
  \mathcal{M} = \{z\in\RR^d|f(z)=0\}
\end{equation}
can be defined solely on basis of the fast dynamics. In cases where
the slow manifold is 1-dimensional (as is the case in all examples
given above), we can identify all points on the slow manifold by
sweeping through the possible values of the (scalar) control parameter
$\mu$. Note that in this case the stable branches of the slow manifold
$\mathcal{M}$ are readily found by relaxing the fast dynamics
\begin{equation}
  \label{eq:fast-dynamics}
  \dot z = f(z)\,,
\end{equation}
i.e. relaxing equation~\eqref{eq:fast-dynamics} for long times until
convergence. The same is not true for the unstable branches,
bifurcation points, etc, which cannot be obtained by this
procedure. 

Instead, we propose a scheme based on the string method
\cite{e-ren-vanden-eijnden:2002, e-ren-vanden-eijnden:2007}. In
simplified terms, the string method computes the heteroclinic orbit
between stable fixed points by relaxing an elastic ``string'' between
them. Applied to the problem at hand of computing $\mathcal{M}$ along
$\mu$, choose a fixed value $\mu_0$ and compute two fixed points $z_A$
and $z_B$ of the fast dynamics~\eqref{eq:fast-dynamics} in the
subspace $\mu=\mu_c$. Since the fast dynamics leave the control
parameter $\mu$ invariant, it is sufficient to choose initial
conditions which fulfill $\mu=\mu_c$. Of course, $z_{i}, i=A,B$ is the
intersection point of $\mathcal{M}$ with the subset $\mu=\mu_c$, and
consequently $z_{i}\in\mathcal{M}$. Now, consider a family of
configurations $\bar z(t,s)\in\RR^d$, with $s\in[0,1]$ (the ``string''), with
$\bar z(t,0)=z_A$ and $\bar z(t,1)=z_B$, in this subspace, connecting the
two fixed points. Relax this family of configurations according to
\begin{equation*}
  \partial_t \bar z(t,s) = (1-\hat t\otimes \hat t)f(\bar z(t,s))\qquad\forall\,s\,\in\,[0,1]\,,
\end{equation*}
where $\hat t = \partial_s \bar z(t,s)/|\partial_s \bar z(t,s)|$ is
the unit tangent vector along the string and therefore
$(1-\hat t\otimes \hat t)$ projects onto the component normal to the
string. After convergence, the resulting string will necessarily have
that $\partial_s \bar z(t,s)$ is parallel to $f(z(t,s))$ everywhere,
and thus forms the heteroclinic orbit between $z_A$ and $z_B$. Since
this heteroclinic orbit necessarily contains a saddle point $z_S$,
which is easily found by identifying the value of $s$ for which
$f(z(t,s))=0$, we have identified the unstable branch of $\mathcal{M}$
in the subspace $\mu=\mu_c$. Repeating the procedure for all values of
$\mu_c$, we can line out the complete slow manifold. Bifurcation
points are identified by points where stable fixed points merge with
saddles. In the case of the subcritical pitchfork bifurcation, we can
even identify all 5 branches of $\mathcal{M}$ in a single string. The
slow manifolds of all infinite dimensional applications above have
been computed by this method. In the finite dimensional case, the slow
manifold is often available analytically.

\section*{References}

\providecommand{\newblock}{}

%\bibliographystyle{iopart-num}
%\bibliography{bib}

\end{document}